\documentclass[twocolumn,amssymb,secnumarabic, prl, aps, nofootinbib]{revtex4-1}

\usepackage{graphicx}
\usepackage{amsmath}
\usepackage{braket}
\usepackage[colorlinks,linkcolor=black,citecolor=black,urlcolor=black,filecolor=black]{hyperref}

\usepackage[export]{adjustbox}

\usepackage{tikz}
\usetikzlibrary{decorations.pathreplacing}
\usetikzlibrary{arrows,arrows.meta,calc,shapes.geometric,shapes.misc}
\tikzset{
	>=stealth',
	help lines/.style={dashed, thick},
	important line/.style={thick},
	connection/.style={thick, dotted},
}

\begin{document}

\title{Probing Topological Spin Liquids on a Programmable Quantum Simulator}

\author
{G. Semeghini$^{1}$,  H. Levine$^{1}$, A. Keesling$^{1}$, S. Ebadi$^{1}$, T. T. Wang$^{1}$, D. Bluvstein$^{1}$, R. Verresen$^{1}$, H. Pichler$^{2,3}$, \\ M. Kalinowski$^{1}$, R. Samajdar$^{1}$, A. Omran$^{1,4}$, S. Sachdev$^{1}$, A. Vishwanath$^{1}$, M. Greiner$^{1}$, V. Vuleti\'{c}$^{5}$, M. D. Lukin$^{1}$}
\affiliation{$^1$Department of Physics, Harvard University, Cambridge, MA 02138, USA \\ $^2$Institute for Theoretical Physics, University of Innsbruck, Innsbruck A-6020, Austria \\ $^3$Institute for Quantum Optics and Quantum Information, Austrian Academy of Sciences, Innsbruck A-6020, Austria \\ $^4$ QuEra Computing Inc., Boston, MA 02135, USA \\ $^5$ Department of Physics and Research Laboratory of Electronics, Massachusetts Institute of Technology, Cambridge, MA 02139, USA}

\begin{abstract}
    Quantum spin liquids, exotic phases of matter with topological order, have been a major focus of explorations in physical science for the past several decades. Such phases feature long-range quantum entanglement that can potentially be exploited to realize robust quantum computation. We use a 219-atom programmable quantum simulator to probe quantum spin liquid states. In our approach, arrays of atoms are placed on the links of a kagome lattice and evolution under Rydberg blockade creates frustrated quantum states with no local order. The onset of a quantum spin liquid phase of the paradigmatic toric code type is detected by evaluating topological string operators that provide direct signatures of topological order and quantum correlations. Its properties are further revealed by using an atom array with nontrivial topology, representing a first step towards topological encoding. Our observations enable the controlled experimental exploration of topological quantum matter and  protected quantum information processing.
\end{abstract}

\maketitle 

Motivated by visionary theoretical work carried out over the past five decades, a broad  search is currently underway 
to identify signatures of quantum spin liquids (QSL) in novel materials~\cite{WenReview2017,SachdevReview2018}.
Moreover, inspired by the  intriguing predictions of quantum information theory~\cite{Kitaev2003fault}, techniques to engineer such systems for topological protection of quantum information are being actively explored \cite{Nayak2008}.
Systems with frustration \cite{savary2016quantum} caused by the lattice geometry or long-range interactions constitute a promising avenue in the search for QSLs. In particular, such systems can be used to implement a class of so-called dimer models \cite{Rokhsar88,Read91,SachdevPRB1992,Moessner01,Misguich02}, which 
are among the most promising candidates to host quantum spin liquid states. However, realizing and probing such states is challenging since they 
are often surrounded by other competing phases.
Moreover, in contrast to topological systems involving time-reversal symmetry breaking, such as in the fractional quantum Hall effect \cite{Halperin20}, these states cannot be easily probed via, e.g., quantized conductance or edge states. Instead, to diagnose spin liquid phases, it is essential to access nonlocal observables, such as topological string operators \cite{WenReview2017,SachdevReview2018}.    
While some indications of QSL phases in correlated materials have been previously reported \cite{YoungLee2012,Imai2015}, thus far, these exotic states of matter have evaded direct experimental detection.

Programmable quantum simulators are well suited for the controlled exploration of these strongly correlated quantum phases \cite{Gross2017,weimer_rydberg_2010,Hermele2009,Yao2013,Glaetzle14,Celi20,BrowaeysSSH}. In particular, recent work showed that various phases of quantum dimer models can be efficiently implemented using Rydberg atom arrays \cite{Rhine2020a} and that a dimer spin liquid state of the toric code type could be potentially created in a specific frustrated lattice \cite{Ruben2020}. We note that toric code states have been dynamically created in small systems using quantum circuits \cite{JWPanPRL2018,Wallraff2020}. However, some of the key properties, 
such as topological robustness, are challenging to realize in such systems.  
Spin liquids have also been explored using quantum annealers, but the lack of coherence in these systems has precluded the observation of quantum features \cite{Zhou20}. 

\begin{figure*} 
\includegraphics[width=150mm]{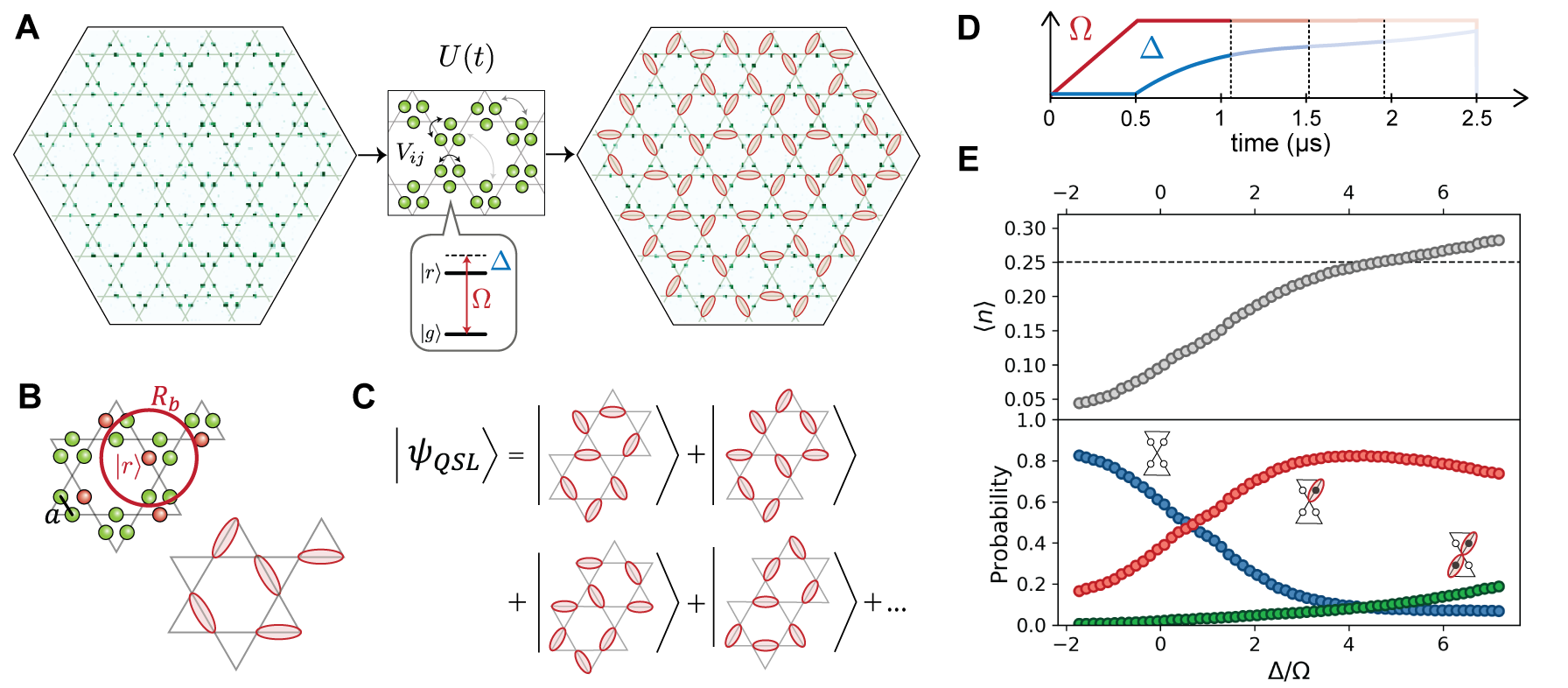}
\caption{\textbf{Dimer model in Rydberg atoms arrays.} (\textbf{A}) Fluorescence image of 219 atoms arranged on the links of a kagome lattice. The atoms, initially in the ground state $|g\rangle$, evolve according to the many-body dynamics $U(t)$. 
The final state of the atoms is determined via fluorescence imaging of ground state atoms. Rydberg atoms are
marked  with red dimers on the bonds of the kagome lattice. (\textbf{B})  We adjust the blockade radius to $R_b/a=2.4$, by choosing $\Omega=2\pi\times1.4$ MHz and $a=3.9~\mu$m,  
such that all six nearest neighbors of an atom in $|r\rangle$ are within the blockade radius $R_b$. 
A state consistent with the Rydberg blockade at maximal filling can then be viewed as a dimer covering of the kagome lattice, where each vertex is touched by exactly one dimer. (\textbf{C}) The quantum spin liquid state corresponds to a coherent superposition of exponentially many dimer coverings. (\textbf{D}) Detuning $\Delta(t)$ and Rabi frequency $\Omega(t)$ used for quasi-adiabatic state preparation. (\textbf{E}) (Top) Average density of Rydberg excitations $\langle n \rangle$ in the bulk of the system, excluding the outer three layers \cite{Supplement}. (Bottom) Probabilities of 
empty vertices in the bulk (monomers), vertices attached to a single dimer, or to double dimers (weakly violating  blockade). After $\Delta/\Omega \sim 3$, the system reaches $\sim 1/4$ filling, where most vertices are attached to a single dimer, consistent with an approximate dimer phase.
}
\label{fig1}
\end{figure*}

\noindent \textbf
{Dimer Models in Rydberg Atom Arrays.}
The key idea of our approach is based on a correspondence \cite{Ruben2020} between Rydberg atoms placed on the links of a kagome lattice (or equivalently the {\em sites} of a ruby lattice), as shown in Fig.~\ref{fig1}A,  and dimer models on the kagome lattice \cite{SachdevPRB1992,Misguich02}.
The Rydberg excitations can be viewed as ``dimer bonds'' connecting the two adjacent vertices of the lattice (Fig.~\ref{fig1}B). Due to the Rydberg blockade~\cite{Saffman2010}, strong and properly tuned interactions constrain the density of excitations such that each vertex is touched by a maximum of one dimer. At 1/4 filling, each vertex is touched by exactly one dimer, resulting in a perfect dimer covering of the lattice.
Smaller filling fractions result in a finite density of vertices with no proximal dimers, which are referred to as {\it monomers}. A quantum spin liquid can emerge within this dimer-monomer model close to 1/4~filling \cite{Ruben2020}, and can be viewed as a coherent superposition of exponentially many degenerate dimer coverings with a
small admixture of monomers
~\cite{Misguich02}
 (Fig.~\ref{fig1}C).
This corresponds to the resonating valence bond (RVB) state \cite{Anderson73,Rokhsar88}, predicted long ago but so far still unobserved in any experimental system. 

\begin{figure}
\includegraphics[width=\columnwidth]{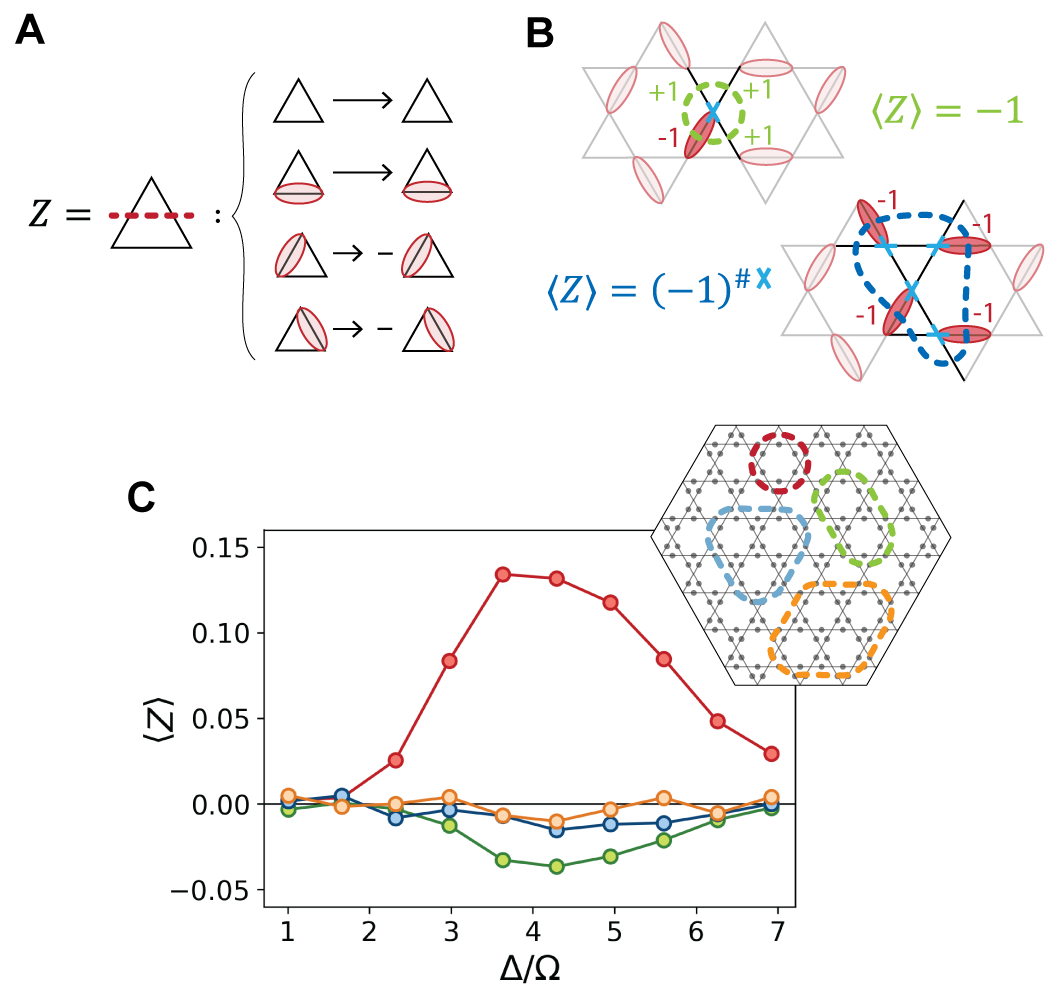}
\caption{\textbf{Detecting a dimer phase via diagonal string operator.} (\textbf{A}) The $Z$ string operator measures the parity of dimers along a string. (\textbf{B}) A perfect dimer covering always has exactly one dimer touching each vertex of the array, so that $\langle Z \rangle=-1$ around a single vertex and $\langle Z \rangle = (-1)^{\# \textnormal{enclosed vertices}}$ for larger loops. (\textbf{C}) $Z$ parity measurements following the quasi-adiabatic sweep of Fig.~\ref{fig1}D, with the addition of a 200~ns ramp-down of $\Omega$ at the end to optimize preparation. At different endpoints of the sweep and for different loop sizes (inset), we measure a finite $\langle Z \rangle$, consistent with an approximate dimer phase.}
\label{fig2}
\end{figure}

To create and study such states experimentally, we utilize two-dimensional arrays of 219 $^{87}$Rb atoms individually trapped in optical tweezers \cite{Ebadi2020, Scholl2020} and positioned on the links of a kagome lattice, as shown in Fig.~\ref{fig1}A.
The atoms are initialized in an electronic ground state $|g \rangle$ and coupled to a Rydberg state $|r\rangle$ via a two-photon optical transition with Rabi frequency $\Omega$. The atoms in the Rydberg state $|r\rangle$ interact via a strong van der Waals potential $V = V_0/d^6$, with $d$ the interatomic distance. 
This strong interaction prevents the simultaneous excitation of two atoms within a blockade radius $R_b = (V_0/\Omega)^{1/6}$ \cite{Saffman2010}. We adjust the lattice spacing $a$ and the Rabi frequency $\Omega$ such that, for each atom in $|r\rangle$, its six nearest neighbors are all within the blockade radius (Fig.~\ref{fig1}B), resulting in a maximum filling fraction of 1/4.
The resulting dynamics corresponds to unitary evolution $U(t)$ governed by the Hamiltonian
\begin{equation}
    \frac{H}{\hbar} = \frac{\Omega(t)}{2}\sum_i \sigma^x_i - \Delta(t)\sum_i n_i + \sum_{i<j} V_{ij}n_i n_j
\label{ryd_ham}
\end{equation}
where $\hbar$ is the reduced Planck constant, $n_i = |r_i\rangle \langle r_i |$ is the Rydberg state occupation at site $i$, $\sigma^x_i = |g_i\rangle \langle r_i| + |r_i\rangle \langle g_i|$ 
and $\Delta(t)$ is the time-dependent two-photon detuning. 
After the evolution, the state is analyzed by projective readout of ground state atoms 
(Fig.~\ref{fig1}A, right panel) \cite{Ebadi2020}. 

To explore  many-body phases in this system, we utilize quasi-adiabatic evolution, in which we slowly  
turn on the Rydberg coupling $\Omega$ 
and subsequently change the detuning $\Delta$ from negative to positive values 
using a cubic frequency sweep over about 2 $\mu$s (Fig.~\ref{fig1}D).
We stop the cubic sweep at different endpoints and first 
measure the density of Rydberg excitations $\langle  n \rangle$. 
Away from the array boundaries (which result in edge effects permeating just two layers into the bulk), we observe that the average density of Rydberg atoms is uniform across the array (see Fig.~S3 and \cite{Supplement}). 
Focusing on the bulk density, we find that for $\Delta/\Omega \gtrsim 3$, the system reaches the desired filling fraction $\langle n \rangle \sim 1/4$ (Fig.~\ref{fig1}E, top panel). The resulting state does not have any obvious spatial order (Fig.~\ref{fig1}A) and appears as a different
configuration of Rydberg atoms in each experimental repetition (see Fig.~S4 and \cite{Supplement}). 
From the single-shot images, we evaluate the probability for each vertex of the kagome lattice to be attached to: one dimer (as in a perfect dimer covering), zero dimers (i.e. a monomer), or two dimers (representing weak blockade violations). Around $\Delta/\Omega\sim 4$ we observe an approximate plateau where $\sim 80\%$ of the vertices are connected to a single dimer (Fig.~\ref{fig1}E), indicating an approximate dimer covering.

\noindent \textbf{Measuring topological string operators.}
A defining property of a phase with topological order is that it cannot be probed locally. Hence, to investigate the possible presence of a QSL state, it is essential to measure topological string operators, analogous to those used in the toric code model \cite{Kitaev2003fault}. For the present model, there are two such string operators, the first of which characterizes the effective dimer description, while the second probes quantum coherence between dimer states \cite{Ruben2020}.
We first focus on the diagonal operator $Z=\prod_{i \in S}
\sigma_i^z$, with $\sigma_i^z = 1- 2n_i$,
that measures the parity of Rydberg atoms along a string $S$ perpendicular to the bonds of the kagome lattice (Fig.~\ref{fig2}A).
For the smallest closed $Z$ loop, which encloses a single vertex of the kagome lattice, $\langle Z \rangle = -1$ for any perfect dimer covering.
Larger loops can be decomposed into a product of small loops around all the enclosed vertices, resulting in  
$\langle Z \rangle = (-1)^{\textnormal{\# enclosed vertices}}$ (Fig.~\ref{fig2}B).
Note that the presence of 
monomers or double-dimers reduces the effective contribution of each vertex, resulting in a reduced $\langle Z \rangle$. 

To measure $\langle Z \rangle$ for different loops (Fig.~\ref{fig2}C), we evaluate the string observables directly from single-shot images,  
averaging over many experimental repetitions and over all loops of the same shape in the bulk of the lattice \cite{Supplement}. 
In the range of detunings where $\langle n \rangle \sim 1/4$, we clearly observe the 
emergence of 
a finite $\langle Z \rangle$ for all loops, with the sign matching the parity of enclosed vertices, as expected for dimer states (Fig.~\ref{fig2}B). The measured values are generally $|\langle Z \rangle|<1$ and decrease with the loop size, suggesting the presence of a finite density of defects, as discussed below. Nevertheless, these observations indicate that the state we prepare is consistent with an approximate dimer phase.

\begin{figure*}
\includegraphics[width=0.95\textwidth]{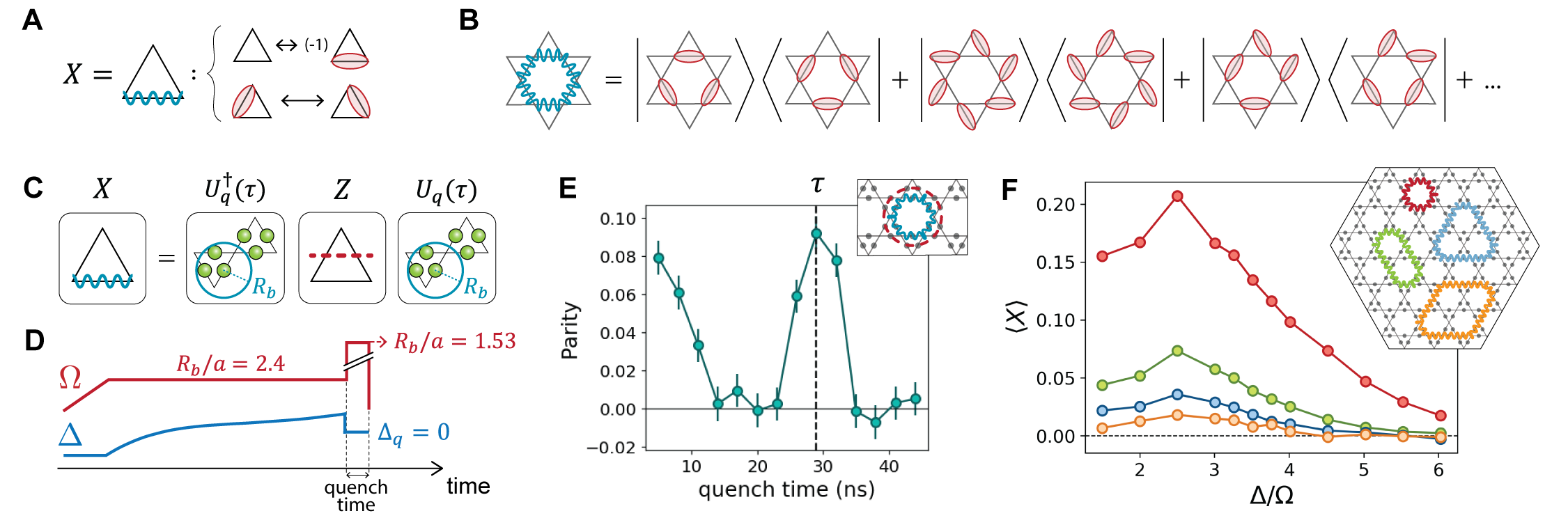}
\caption{\textbf{Probing  coherence between dimer states via off-diagonal string operator.} (\textbf{A})  Definition of $X$ string operator on a single triangle of the kagome lattice. (\textbf{B}) On any closed loop, the $X$ operator maps any dimer covering into another valid dimer covering, such that $\langle X \rangle$ measures the coherence between pairs of dimer configurations. (\textbf{C}) The $X$ operator is 
measured 
by evolving the initial state under  Hamiltonian (eq.~\eqref{ryd_ham}) with $\Delta=0$ and reduced blockade radius to encompass only atoms within each individual triangle, implementing a basis rotation that maps $X$ into $Z$. (\textbf{D}) In the experiment, after the  state preparation, we set the laser detuning to $\Delta_q=0$ and we increase $\Omega$ to $2\pi\times 20$~MHz to reach $R_b/a=1.53$. (\textbf{E}) By measuring the $Z$ parity on the dual string (red) of a target $X$ loop (blue) after a variable quench time, we identify the time $\tau$ for which the mapping in (\textbf{C}) is implemented. (\textbf{F}) We measure $\langle X \rangle$ for different final detunings of the cubic sweep and for different loop sizes (inset), and find that the prepared state  has long-range coherence that extends over a large fraction of the array \cite{Supplement}.
}
\label{fig3}
\end{figure*}

We next explore quantum coherence properties of the prepared state.
To this end, we consider the off-diagonal  
$X$ operator, which acts on strings along the bonds of the kagome lattice. It is defined in Fig.~\ref{fig3}A by its action on a single triangle \cite{Ruben2020}.
Applying $X$ on any closed string maps a dimer covering to another valid dimer covering (see e.g.  Fig.~\ref{fig3}B for a loop around a single hexagon).
A finite expectation value for $X$ therefore implies that the state contains a coherent superposition of one or more pairs of dimer states coupled by that specific loop,  a prerequisite for a {\em quantum} spin liquid. 
The measurement of $X$ can be implemented by 
performing a collective basis rotation \cite{Ruben2020} illustrated in Fig.~\ref{fig3}C. 
This rotation is implemented by time-evolution 
under the Rydberg Hamiltonian (eq.~\eqref{ryd_ham}) with $\Delta=0$ and reduced blockade radius $R_b/a=1.53$, such that only the atoms within the same triangle are subject to the Rydberg blockade constraint. Under these conditions, it is sufficient to consider the evolution of individual triangles separately, where each triangle can be described as a 4-level system ($\includegraphics[valign=c,width=20mm]{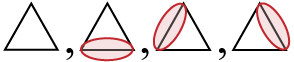}$). Within this subspace, after a time $\tau=4\pi/(3\Omega\sqrt{3})$, the collective 3-atom dynamics realizes a unitary $U_q$ which  implements the basis rotation that transforms an $X$ string into a dual $Z$ string \cite{Supplement}. 

Experimentally, the basis rotation is implemented following the state preparation
by quenching the laser detuning to $\Delta_q=0$ and increasing the laser intensity by a factor of $\sim 200$ to reduce the blockade radius to $R_b/a=1.53$ (Fig.~\ref{fig3}D and \cite{Supplement}). 
We  calibrate $\tau$ by preparing the state at $\Delta/\Omega=4$ and
evolving under the quench Hamiltonian 
for a variable time. We measure the parity of a $Z$ string that is dual to a target $X$ loop, and observe a sharp revival of the parity signal at $\tau\sim 30$~ns (Fig.~\ref{fig3}E) \cite{Ruben2020}.
Fixing the quench time $\tau$, we measure $\langle X \rangle$ for different values of the detuning $\Delta$ at the end of the cubic sweep (Fig.~\ref{fig3}F) and  observe a finite $X$ parity signal for loops that extend over a large fraction of the array. We emphasize that, in light of experimental imperfections \cite{Supplement}, the observation of finite parities for string observables of up to 28 atoms
within $\mu$s-long experiments is rather remarkable. These observations clearly indicate the presence of long-range coherence in the prepared state. 

\begin{figure}
\includegraphics[width=\columnwidth]{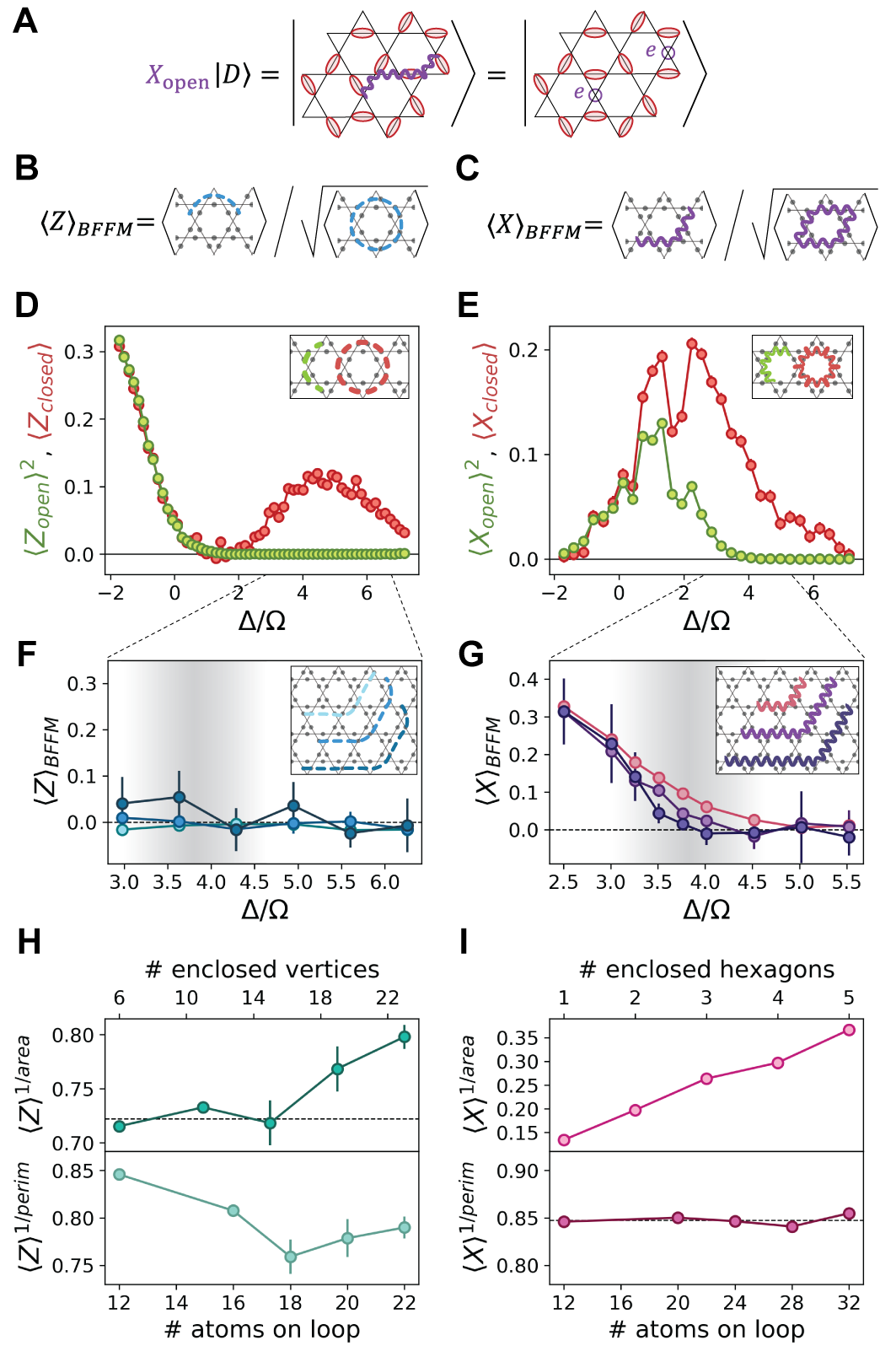}
\caption{\textbf{String order parameters and quasiparticle excitations.} (\textbf{A}) An open string operator $X_{\textrm{open}}$ acting on a dimer state $|D\rangle$ creates two monomers ($e$-anyons) at its endpoints (see Fig.~S9 for $m$-anyons). (\textbf{B,C}) Definition of the string order parameters $\langle Z \rangle_{\textnormal{BFFM}}$ and $\langle X \rangle_{\textnormal{BFFM}}$. (\textbf{D})
Comparison between $\langle Z_{\textnormal{closed}} \rangle$ and $\langle Z_{\textnormal{open}} \rangle^2$ measured on the strings shown in the inset. The expectation value shown for the open string is squared to account for the different length of the strings.
(\textbf{E}) Analogous comparison for $X$. (\textbf{F,G}) Zooming in on the range  with finite closed loop parities 
we measure the BFFM order parameters for different open strings (insets). We find that $\langle Z \rangle_{\textnormal{BFFM}}$ is consistent with zero on the entire range of $\Delta$, while $\langle X \rangle_{\textnormal{BFFM}}$ vanishes for $\Delta/\Omega\gtrsim 3.3$, allowing us 
to identify a range of detunings consistent with the onset of a QSL phase (shaded area). (\textbf{H}) Rescaled parities $\langle Z \rangle^{1/\textnormal{area}}$ and $\langle Z \rangle^{1/\textnormal{perim}}$ evaluated for $\Delta/\Omega=3.6$, where area and perimeter are defined as the number of vertices enclosed by the loop and the number of atoms on the loop, respectively. For small loops, $Z$ scales with an area law, while it deviates from this behavior for larger loops, converging towards a perimeter law. (\textbf{I}) $\langle X \rangle^{1/\textnormal{area}}$ (the area, in this case, is the number of enclosed hexagons) and $\langle X \rangle^{1/\textnormal{perim}}$ evaluated for $\Delta/\Omega=3.5$, indicating an excellent agreement with a perimeter-law scaling.}
\label{fig4}
\end{figure}

\noindent \textbf{Probing spin liquid properties.}
The study of closed string operators shows that we prepare an approximate dimer phase with quantum coherence between dimer coverings.
While these closed loops are indicative of topological order, it is important to compare their properties to those of open strings to distinguish topological effects from trivial ordering---the former being sensitive to the topology of the loop \cite{BF1983,FM1983,Gregor11}.
This comparison is shown in  Fig.~\ref{fig4}D,E, indicating several distinct regimes.  
For small $\Delta$, we find that both $Z$ and $X$ loop parities factorize into the product of the parities on the half-loop open strings---in particular,
the finite $\langle Z \rangle$ is a trivial result of the low density of Rydberg excitations. In contrast, loop parities no longer factorize 
in the dimer phase ($3\lesssim \Delta/\Omega \lesssim 5$). Instead, the expectation values for both open string operators vanish in the dimer phase, indicating the nontrivial nature of the correlations measured by the closed loops (see also \cite{Supplement}). More specifically, 
topological ordering in the dimer-monomer model can break down either due to a high density of monomers, corresponding to the trivial disordered phase at small $\Delta/\Omega$, or due to the lack of long-range resonances, corresponding to a valence bond solid (VBS) \cite{Ruben2020}.
Open $Z$ and $X$ strings distinguish the target QSL phase from these proximal phases:  when normalized according to the definition from Bricmont, Fr{\"o}lich, Fredenhagen and Marcu \cite{BF1983,FM1983} (BFFM) (Fig.~\ref{fig4}B,C), these open strings can be considered as order parameters for the QSL. In particular, open $Z$ strings have a finite expectation value when the dimers form an ordered spatial arrangement, as in the VBS phase. At the same time, open $X$ strings create pairs of monomers at their endpoints (Fig.~\ref{fig4}A), so a finite $\langle X \rangle$ can be achieved in the trivial phase where there is a high density of monomers. 
Therefore, the QSL can be identified as the unique phase where both order parameters vanish for long strings \cite{Ruben2020}.

Figures~\ref{fig4}F,G show the measured values of 
these order parameters. 
We find that $\langle Z \rangle_{\textnormal{BFFM}}$
is compatible with zero on the entire range of $\Delta/\Omega$ where we observed a finite $Z$ parity on closed loops, indicating the absence of a VBS phase (Fig.~\ref{fig4}F), consistent with our analysis of density-density correlations (Fig.~S5 and \cite{Supplement}).
At the same time, $\langle X\rangle_{\textnormal{BFFM}}$ converges towards zero on the longest strings for $\Delta/\Omega \gtrsim 3.3$ (Fig.~\ref{fig4}G), indicating a transition out of the disordered phase. By combining these two measurements with the regions of non-vanishing parity for the closed $Z$ and $X$ loops (Figs.~\ref{fig2},\ref{fig3}), we conclude that for $3.3 \lesssim \Delta/\Omega \lesssim 4.5$ our results constitute a direct detection of the onset of a quantum spin liquid phase (shaded area in Fig.~\ref{fig4}F,G). 

The measurements of the closed loop operators in Fig.~\ref{fig2},\ref{fig3} show that $|\langle Z \rangle|, |\langle X \rangle| <1$ and that the amplitude of the signal decreases with the loop size, which results from a finite density of quasiparticle excitations.
Specifically, defects in the dimer covering such as monomers and double-dimers can be interpreted as {\it electric} ($e$) anyons in the language of lattice gauge theory \cite{Ruben2020}.   
Since the presence of a defect inside a closed loop changes the sign of $Z$, the parity on the loop is reduced according to the number of enclosed $e$-anyons as
$|\langle Z \rangle| = |\langle(-1)^{\# \textnormal{enclosed e-anyons}} \rangle|$.
The average number of defects inside a loop is expected to scale with the number of enclosed vertices, i.e. with the {\it area} of the loop, and indeed
we observe an approximate { \it area-law} scaling of $|\langle Z \rangle|$ for small loop sizes (Fig.~\ref{fig4}H).
However, for larger loops we notice a deviation towards a {\it perimeter-law} scaling,
which can emerge if pairs of anyons are correlated over a characteristic length scale smaller than the loop size (see \cite{Supplement} for a discussion of the expected scaling).
Pairs of correlated anyons which are both inside the loop 
do not change its parity since their contributions cancel out; they only affect $\langle Z \rangle$ when they sit across the loop, leading to a scaling with the length of the perimeter. These 
pairs  
can be viewed as resulting from the application of $X$ string operators to a dimer covering (Fig.~\ref{fig4}A), originating, e.g., from  virtual excitations in the dimer-monomer model \cite{Supplement} or from errors due to state preparation and detection. Note that state preparation with larger Rabi frequency (improved adiabaticity) results in larger $Z$ parity signals and reduced $e$-anyon density (see Fig.~S7). 

A second type of quasiparticle excitation that could arise in this model is the so-called {\it magnetic} ($m$) anyon. Analogous to $e$-anyons which live at the endpoints of open $X$ strings (Fig.~\ref{fig4}A), $m$-anyons are created by open $Z$ strings and they correspond to phase errors between dimer coverings (Fig.~S9 and \cite{Supplement}). These excitations cannot be directly identified from individual snapshots, but they are detected by the measurement of closed $X$ loop operators. The remarkable perimeter law scaling observed in Fig.~\ref{fig4}I indicates that $m$-anyons only appear in pairs with short correlation lengths \cite{Supplement}.
These observations highlight the prospects for using topological string operators to detect and probe quasiparticle excitations in the system.

\begin{figure*}
\includegraphics[width=105mm]{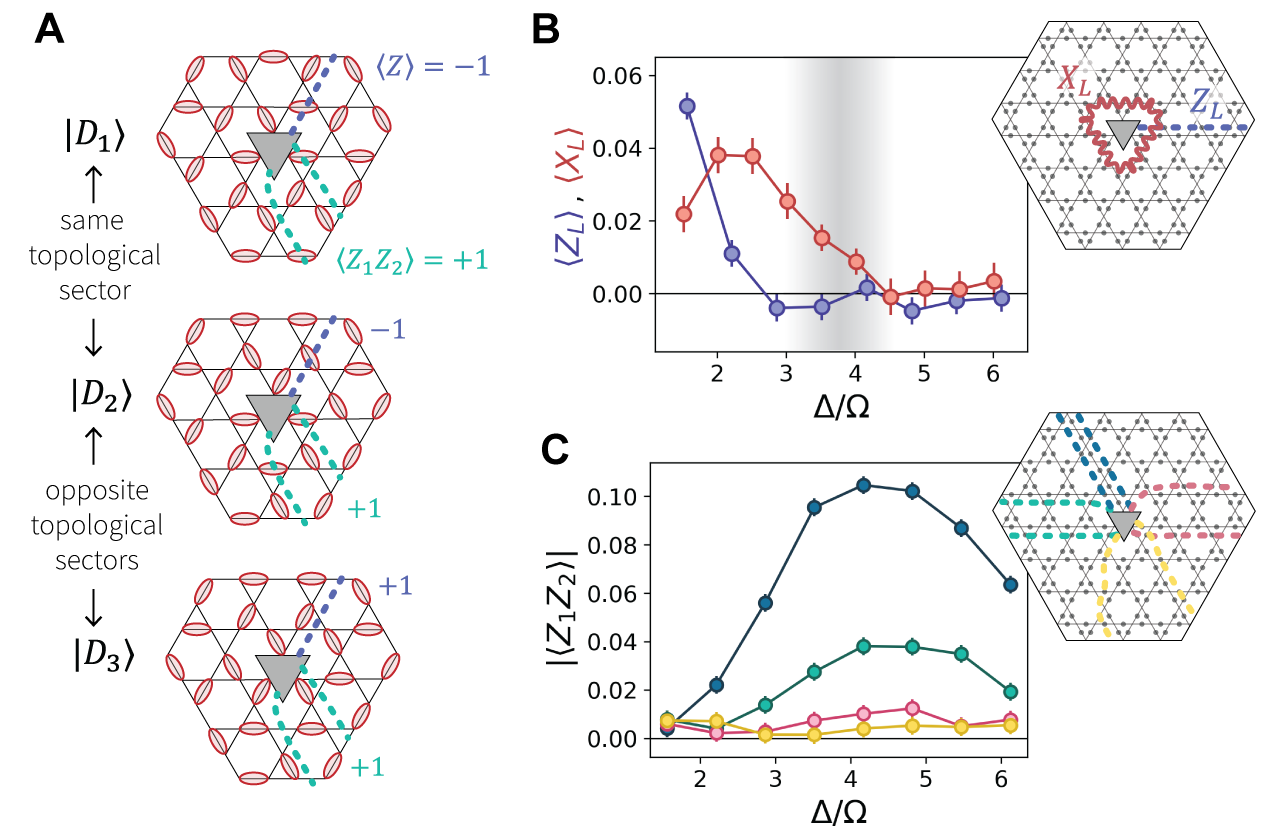}
\caption{\textbf{Topological properties in array with a hole.} (\textbf{A}) A lattice with nontrivial topology is obtained by removing three atoms at the center to create a small hole. The dimer states can be divided into two distinct topological sectors 0 and 1. 
$Z$ strings connecting the hole to the boundary always have a well-defined expectation value within each sector and opposite sign between the two sectors, while the correlations between two such strings $Z_1 Z_2$ are identical for both sectors. (\textbf{B}) Measured expectation values for the operators $Z_L$ and $X_L$ defined in the inset, indicate that in the QSL region (shaded area)
we prepare a superposition state of the two topological sectors ($\langle Z_L \rangle=0$) with 
a finite overlap with the $|+\rangle$ state ($\langle X_L \rangle>0$). (\textbf{C}) Finite expectation values for the correlations between pairs of hole-to-boundary $Z$ strings (inset), consistent with (\textbf{A}).}
\label{fig5}
\end{figure*}

\noindent \textbf{Towards a topological qubit.}
To further explore the topological properties of the spin liquid state, we create an atom array with a small hole
by removing three atoms on a central triangle, which creates an effective inner boundary (Fig.~\ref{fig5}). This results in two distinct topological sectors for the dimer coverings, where states belonging to different sectors can be transformed into each other only via large $X$ loops which enclose the hole, constituting
a highly nonlocal process (involving at least a 16-atom resonance) (Fig.~S11). 
We define the logical states $|0_L\rangle$ and $|1_L\rangle$ as the superpositions of all
dimer coverings from the topological sectors 0 and 1, respectively. One can define \cite{Ruben2020} the logical operator $\sigma^z_L$ as proportional to 
any $Z_L$ string operator that connects the hole with the outer boundary, since these have a well-defined eigenvalue $\pm 1$ for all dimer states in the same sector but opposite for the two sectors. The logical $\sigma^x_L$ is instead proportional to $X_L$, which is any $X$ loop around the hole. This operator anticommutes with $Z_L$ and has eigenstates $|+ \rangle \sim (|0_L\rangle + |1_L\rangle)/\sqrt{2}$ and $|- \rangle \sim (|0_L\rangle - |1_L\rangle)/\sqrt{2}$. 

We measure $Z_L$ and $X_L$ on the strings defined in the inset of Fig.~\ref{fig5}B, following the same quasi-adiabatic preparation as in Fig.~\ref{fig1}D.
We find that in the range of $\Delta/\Omega$ associated with the onset of a QSL phase, $\langle Z_L \rangle = 0$, and $\langle X_L \rangle > 0$, indicating that the system is in a superposition of the two topological sectors, with 
a finite overlap with the $|+\rangle$ state (Fig.~\ref{fig5}B). To further support this conclusion,  we evaluate correlations $\langle Z_1 Z_2\rangle$ between hole-to-boundary strings, which are expected to have the same expectation values for both topological sectors (Fig.~\ref{fig5}A). 
In agreement with this prediction,  we find that the correlations between different pairs of strings have finite expectation values, with amplitudes decreasing with the distance between the strings (Fig. \ref{fig5}C) due to imperfect state preparation.
These measurements
represent the first steps towards initialization and measurement of a
topological qubit. 
\\

\noindent \textbf{Discussion and outlook.}
Noting that it is not possible to classically simulate quantum dynamics for the full experimental system,
we compare our results with several theoretical approaches. We first note that our observations qualitatively disagree with the  ground state phase diagram obtained from 
density-matrix-renormalization-group (DMRG) \cite{White92,Hauschild18} simulations on infinitely-long cylinders. 
For the largest accessible system sizes, including van der Waals interactions only up to intermediate distances ($\sim 4a$), we find a $\mathbb Z_2$ spin liquid in the ground state. However, unlike in deformed lattices \cite{Ruben2020}, longer-range couplings destabilize the spin liquid in the ground state of the Hamiltonian (eq.~\eqref{ryd_ham}) on the specific ruby lattice used in the experiment, 
leading to a direct first-order transition from the disordered phase to the VBS phase \cite{Supplement}. In contrast, we experimentally observe the onset of the QSL phase in a relatively large parameter range, 
while no signatures of a VBS phase are detected. 

To develop additional insight, we perform time-dependent DMRG calculations \cite{White92,Zaletel15,Hauschild18} simulating the same state preparation protocol as in the experiment on an infinitely-long cylinder with a seven-atom-long circumference \cite{Supplement}. The results of these simulations are in good qualitative agreement with our experimental observations (see Fig.~S17). 
Specifically, similar to the results in Fig.~\ref{fig4}, we find that the region $\Delta/\Omega \sim 3.5–4.5$ hosts nonzero signals for closed $Z$ and $X$ loops which cannot be factorized into open strings, a characteristic fingerprint of spin liquid correlations.
In addition, exact diagonalization studies of a simplified blockade model reveal how the dynamical state preparation creates an approximate equal-weight and equal-phase superposition of many dimer states, instead of the VBS ground state \cite{Supplement}.
We conclude that quasi-adiabatic state preparation occurring over a few microseconds
is insensitive to longer-range couplings and generates states that retain the QSL character \cite{Supplement}. 
While this phenomenon deserves further theoretical studies, these considerations point towards the creation of a novel {\it metastable} state with key characteristic properties of a quantum spin liquid. 

Our experiments offer unprecedented  insights into  elusive topological quantum matter, and open up a number of new directions in which these studies can be extended, including:
improving the robustness of the QSL by using modified lattice geometries and boundaries \cite{Rhine2020a,Ruben2020}, as well as 
optimizing the state preparation to minimize quasiparticle excitations; understanding and mitigating environmental effects associated, e.g., with dephasing and spontaneous emission \cite{Supplement};  
optimizing string operator measurements using quasi-local transformations \cite{HastingsWen2005}, potentially with the help of
quantum algorithms \cite{Cong2019}. 
At the same time, hardware-efficient techniques for  
robust manipulation and braiding of topological qubits can be explored. Furthermore, methods for anyon trapping and annealing can be investigated, with  eventual applications towards fault-tolerant quantum information processing \cite{KitaevPreskill4D}. With improved programmability and control, a broader class of topological quantum matter and lattice gauge theories can be efficiently implemented \cite{CiracLGT2020, Lewenstein2013}, opening the door to their detailed exploration under controlled experimental conditions, and providing a novel route for the design of quantum materials that can supplement exactly solvable models~\cite{Kitaev2003fault, KitaevHoneycomb} and classical numerical methods \cite{White92,Hauschild18}. 
\\\\
\noindent {\it Note added}: During the completion of this manuscript we became aware of related work demonstrating the preparation of toric code states on a 32-qubit superconducting quantum processor \cite{Google2021}.
\\\\
\textbf{Acknowledgments} 
\\
We thank many members of the Harvard AMO community, particularly E. Urbach, S. Dakoulas, and J. Doyle for their efforts enabling  operation of our laboratories during 2020-2021. We thank S. Choi, I. Cong, E. Demler, X. Gao, G. Giudici, W. W. Ho, N. Maskara, K. Najafi, N. Yao and S. Yelin for stimulating discussions. \textbf{Funding:} We acknowledge financial support from the Center for Ultracold Atoms, the National Science Foundation, the U.S. Department of Energy (DE-SC0021013 \& LBNL QSA Center), the Army Research Office, ARO MURI, and the DARPA ONISQ program. G.S. acknowledges support from a fellowship from the Max Planck/Harvard Research Center for Quantum Optics. H.L. acknowledges support from the National Defense Science and Engineering Graduate (NDSEG) fellowship. T.T.W. acknowledges support from Gordon College. D.B. acknowledges support from the NSF Graduate Research Fellowship Program (grant DGE1745303) and The Fannie and John Hertz Foundation. R.V. acknowledges support from the Harvard Quantum Initiative Postdoctoral Fellowship in Science and Engineering. R.V., A.V. and S.S. acknowledge support from the Simons Collaboration on Ultra-Quantum Matter, which is a grant from the Simons Foundation (651440, A.V., S.S.). R.S. and S.S. were supported by the U.S. Department of Energy under Grant DE-SC0019030. The DMRG simulations were performed using the Tensor Network Python (TeNPy) package developed by Johannes Hauschild and Frank Pollmann \cite{Hauschild18}, and they were run on the FASRC Cannon and Odyssey clusters supported by the FAS Division of Science Research Computing Group at Harvard University. \textbf{Author contributions:} G.S., H.L., A.K., S.E., T.T.W., D.B., and A.O. contributed to building the experimental setup, performed the measurements, and analyzed the data. R.V., H.P. and A.V. contributed to developing methods for detecting QSL correlations, performed numerical simulations and contributed to the theoretical interpretation of the results. M.K. and R.S. contributed to the theoretical interpretation of the results. All work was supervised by S.S., M.G., V.V., and M.D.L. All authors discussed the results and contributed to the manuscript. \textbf{Competing interests:} M.G., V.V., and M.D.L. are co-founders and shareholders of QuEra Computing. A.O. is a shareholder of QuEra Computing.

\let\oldaddcontentsline\addcontentsline
\renewcommand{\addcontentsline}[3]{}
\bibliographystyle{Science}
\bibliography{SLbib.bib}

\begin{thebibliography}{10}

\bibitem{WenReview2017}
X.-G. Wen, {\it Rev. Mod. Phys.\/} {\bf 89}, 041004 (2017).

\bibitem{SachdevReview2018}
S.~Sachdev, {\it Reports on Progress in Physics\/} {\bf 82}, 014001 (2018).

\bibitem{Kitaev2003fault}
A.~Kitaev, {\it Annals of Physics\/} {\bf 303}, 2  (2003).

\bibitem{Nayak2008}
C.~Nayak, S.~H. Simon, A.~Stern, M.~Freedman, S.~Das~Sarma, {\it Rev. Mod.
  Phys.\/} {\bf 80}, 1083 (2008).

\bibitem{savary2016quantum}
L.~Savary, L.~Balents, {\it Reports on Progress in Physics\/} {\bf 80}, 016502
  (2016).

\bibitem{Rokhsar88}
D.~S. Rokhsar, S.~A. Kivelson, {\it Phys. Rev. Lett.\/} {\bf 61}, 2376 (1988).

\bibitem{Read91}
N.~Read, S.~Sachdev, {\it Phys. Rev. Lett.\/} {\bf 66}, 1773 (1991).

\bibitem{SachdevPRB1992}
S.~Sachdev, {\it Phys. Rev. B\/} {\bf 45}, 12377 (1992).

\bibitem{Moessner01}
R.~Moessner, S.~L. Sondhi, {\it Phys. Rev. Lett.\/} {\bf 86}, 1881 (2001).

\bibitem{Misguich02}
G.~Misguich, D.~Serban, V.~Pasquier, {\it Phys. Rev. Lett.\/} {\bf 89}, 137202
  (2002).

\bibitem{Halperin20}
B.~I. Halperin, J.~K. Jain, {\it Fractional Quantum Hall Effects\/} (WORLD
  SCIENTIFIC, 2020).

\bibitem{YoungLee2012}
T.-H. Han, {\it et~al.\/}, {\it Nature\/} {\bf 492}, 406 (2012).

\bibitem{Imai2015}
M.~Fu, T.~Imai, T.-H. Han, Y.~S. Lee, {\it Science\/} {\bf 350}, 655 (2015).

\bibitem{Gross2017}
C.~Gross, I.~Bloch, {\it Science\/} {\bf 357}, 995 (2017).

\bibitem{weimer_rydberg_2010}
H.~Weimer, M.~Müller, I.~Lesanovsky, P.~Zoller, H.~P. Büchler, {\it Nature
  Physics\/} {\bf 6}, 382 (2010).

\bibitem{Hermele2009}
M.~Hermele, V.~Gurarie, A.~M. Rey, {\it Phys. Rev. Lett.\/} {\bf 103}, 135301
  (2009).

\bibitem{Yao2013}
N.~Y. Yao, {\it et~al.\/}, {\it Phys. Rev. Lett.\/} {\bf 110}, 185302 (2013).

\bibitem{Glaetzle14}
A.~W. Glaetzle, {\it et~al.\/}, {\it Phys. Rev. X\/} {\bf 4}, 041037 (2014).

\bibitem{Celi20}
A.~Celi, {\it et~al.\/}, {\it Phys. Rev. X\/} {\bf 10}, 021057 (2020).

\bibitem{BrowaeysSSH}
S.~de~L{\'e}s{\'e}leuc, {\it et~al.\/}, {\it Science\/} {\bf 365}, 775 (2019).

\bibitem{Rhine2020a}
R.~Samajdar, W.~W. Ho, H.~Pichler, M.~D. Lukin, S.~Sachdev, {\it Proceedings of
  the National Academy of Sciences\/} {\bf 118}, e2015785118 (2021).

\bibitem{Ruben2020}
R.~Verresen, M.~D. Lukin, A.~Vishwanath, {\it arXiv:2011.12310\/}  (2020).

\bibitem{JWPanPRL2018}
C.~Song, {\it et~al.\/}, {\it Phys. Rev. Lett.\/} {\bf 121}, 030502 (2018).

\bibitem{Wallraff2020}
C.~K. Andersen, {\it et~al.\/}, {\it Nature Physics\/} {\bf 16}, 875 (2020).

\bibitem{Zhou20}
S.~{Zhou}, D.~{Green}, E.~D. {Dahl}, C.~{Chamon}, {\it arXiv e-prints\/} p.
  arXiv:2009.07853 (2020).

\bibitem{Saffman2010}
M.~Saffman, T.~G. Walker, K.~M\o{}lmer, {\it Rev. Mod. Phys.\/} {\bf 82}, 2313
  (2010).

\bibitem{Anderson73}
P.~Anderson, {\it Materials Research Bulletin\/} {\bf 8}, 153 (1973).

\bibitem{Supplement}
Materials and methods are available as supplementary materials.

\bibitem{Ebadi2020}
S.~Ebadi, {\it et~al.\/}, {\it arXiv:2012.12281\/}  (2020).

\bibitem{Scholl2020}
P.~Scholl, {\it et~al.\/}, {\it arXiv:2012.12268\/}  (2020).

\bibitem{BF1983}
J.~Bricmont, J.~Frölich, {\it Physics Letters B\/} {\bf 122}, 73 (1983).

\bibitem{FM1983}
K.~Fredenhagen, M.~Marcu, {\it Communications in Mathematical Physics\/} {\bf
  92}, 81  (1983).

\bibitem{Gregor11}
K.~Gregor, D.~A. Huse, R.~Moessner, S.~L. Sondhi, {\it New Journal of
  Physics\/} {\bf 13}, 025009 (2011).

\bibitem{White92}
S.~R. White, {\it Phys. Rev. Lett.\/} {\bf 69}, 2863 (1992).

\bibitem{Hauschild18}
J.~Hauschild, F.~Pollmann, {\it SciPost Phys. Lect. Notes\/} p.~5 (2018).

\bibitem{Zaletel15}
M.~P. Zaletel, R.~S.~K. Mong, C.~Karrasch, J.~E. Moore, F.~Pollmann, {\it Phys.
  Rev. B\/} {\bf 91}, 165112 (2015).

\bibitem{HastingsWen2005}
M.~B. Hastings, X.-G. Wen, {\it Phys. Rev. B\/} {\bf 72}, 045141 (2005).

\bibitem{Cong2019}
I.~Cong, S.~Choi, M.~D. Lukin, {\it Nature Physics\/} {\bf 15}, 1273 (2019).

\bibitem{KitaevPreskill4D}
E.~Dennis, A.~Kitaev, A.~Landahl, J.~Preskill, {\it Journal of Mathematical
  Physics\/} {\bf 43}, 4452 (2002).

\bibitem{CiracLGT2020}
M.~C. Bañuls, {\it et~al.\/}, {\it The European Physical Journal D\/} {\bf
  74}, 165 (2020).

\bibitem{Lewenstein2013}
L.~Tagliacozzo, A.~Celi, A.~Zamora, M.~Lewenstein, {\it Annals of Physics\/}
  {\bf 330}, 160 (2013).

\bibitem{KitaevHoneycomb}
A.~Kitaev, {\it Annals of Physics\/} {\bf 321}, 2 (2006). January Special
  Issue.

\bibitem{Google2021}
K.~J. Satzinger, {\it et~al.\/}, {\it arXiv:2104.01180\/}  (2021).

\bibitem{rispoli_quantum_2019}
M.~Rispoli, {\it et~al.\/}, {\it Nature\/} {\bf 573}, 385 (2019).

\bibitem{Fradkin79}
E.~Fradkin, S.~H. Shenker, {\it Phys. Rev. D\/} {\bf 19}, 3682 (1979).

\bibitem{Sutherland88}
B.~Sutherland, {\it Phys. Rev. B\/} {\bf 37}, 3786 (1988).

\bibitem{Turner18}
C.~J. Turner, A.~A. Michailidis, D.~A. Abanin, M.~Serbyn,
  Z.~Papi\ifmmode~\acute{c}\else \'{c}\fi{}, {\it Phys. Rev. B\/} {\bf 98},
  155134 (2018).

\bibitem{White93}
S.~R. White, {\it Phys. Rev. B\/} {\bf 48}, 10345 (1993).

\bibitem{Stoudenmire12}
E.~Stoudenmire, S.~R. White, {\it Annual Review of Condensed Matter Physics\/}
  {\bf 3}, 111 (2012).

\bibitem{katoAdiabaticTheoremQuantum1950}
T.~Kato, {\it Journal of the Physical Society of Japan\/} {\bf 5}, 435 (1950).

\bibitem{zurekDynamicsQuantumPhase2005}
W.~H. Zurek, U.~Dorner, P.~Zoller, {\it Physical Review Letters\/} {\bf 95},
  105701 (2005).

\bibitem{dziarmagaDynamicsQuantumPhase2005a}
J.~Dziarmaga, {\it Physical Review Letters\/} {\bf 95}, 245701 (2005).

\bibitem{polkovnikovUniversalAdiabaticDynamics2005}
A.~Polkovnikov, {\it Physical Review B\/} {\bf 72}, 161201 (2005).

\end{thebibliography}
\let\addcontentsline\oldaddcontentsline

\clearpage
\onecolumngrid
\begin{center}
    \textbf{\Large Supplementary Materials}
\end{center}
\normalsize

\setcounter{equation}{0}
\setcounter{figure}{0}
\setcounter{table}{0}
\makeatletter
\renewcommand{\theequation}{S\arabic{equation}}
\renewcommand{\thefigure}{S\arabic{figure}}
\setlength\tabcolsep{10pt}
\setcounter{secnumdepth}{2}
\renewcommand\thesection{\arabic{section}}

\newcommand\numberthis{\addtocounter{equation}{1}\tag{\theequation}}
\newcommand{\insertimage}[1]{\includegraphics[valign=c,width=0.04\columnwidth]{#1}}

\tableofcontents

\section{Experimental system \label{Sec:expt}}
Our experiments make use of the second generation of the atom array setup, described previously in \cite{Ebadi2020}. In our experiments, atoms are excited to Rydberg states using a two-photon excitation scheme, consisting of a 420~nm laser from the ground state $5S_{1/2}$ to the intermediate state $6P_{3/2}$, and a 1013~nm laser from the intermediate state to the Rydberg state $70S_{1/2}$. Details of both laser systems are presented in Ref.~\cite{Ebadi2020}.

\begin{figure}[hb!]
\includegraphics[width=3.5in]{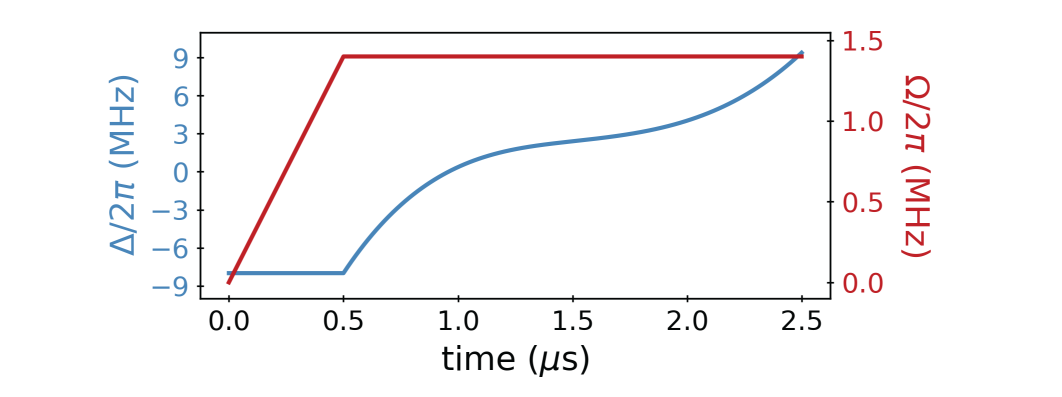}
    \caption{\textbf{Quasi-adiabatic state preparation.} $\Omega(t)$ and $\Delta(t)$ used for state preparation. To probe the phase diagram at different $\Delta$, we stop the cubic sweep at different endpoints and correspondingly turn off $\Omega$.}
\label{fig:CubicSweep}
\end{figure}

In the present work, we tune the lasers to have a detuning of $\delta = 2\pi \times -450$~MHz from the intermediate $6P_{3/2}$ state, where the 420~nm laser is red-detuned from the intermediate state. The 1013~nm laser is always applied at maximum optical power ($\sim 3~$W total on the atoms), and results in a single-photon Rabi frequency $\Omega_\text{1013} = 2\pi \times 50$~MHz.
The 420~nm laser power varies depending on the protocol. During the quasi-adiabatic preparation of the dimer phase, we apply the 420~nm light at low power, which reduces the two-photon Rabi frequency and therefore increases the blockade radius to the target $R_b/a = 2.4$. This low power setting consists of a total of $\sim 0.5$~mW on the atoms, with a single-photon Rabi frequency $\Omega_\text{420} = 2\pi \times 25~$MHz. During the quasi-adiabatic preparation, we therefore have a two-photon Rabi frequency of $\Omega = \Omega_\text{420} \Omega_\text{1013} / 2\delta = 2\pi \times 1.4~$MHz (details of $\Omega(t)$ and $\Delta(t)$ used for state preparation are reported in Fig.~\ref{fig:CubicSweep}).
Under these conditions, we estimate the rate of off-resonant scattering from $\ket{g}$ due to the 420~nm laser to be $\sim 1 / (150~\mu$s$)$, and the decay rate of $\ket{r}$ to be $1 / (80~\mu$s$)$ (including radiative decay, blackbody stimulated transitions, and off-resonant scattering from the 1013~nm laser). 
State detection fidelity for both ground state and Rydberg atoms is $99\%$ \cite{Ebadi2020}.

To measure the $X$ operator, following the dimer phase preparation, we apply short quenches at significantly higher blue power. This high power setting consists of 
a maximum power of $\sim 100~$mW on the atoms, corresponding to a single-photon Rabi frequency $\Omega_\text{420} = 2\pi \times 360$~MHz. The corresponding two-photon Rabi frequency is $\Omega = 2\pi \times 20~$MHz, and $R_b / a = 1.53$. In this configuration, the 420~nm laser introduces a substantially larger light shift on the Rydberg transition of $2\pi \times 36~$MHz. To avoid systematic offsets in the effective detuning from resonance, we separately calibrate the resonance condition at both low power and high power. The 420~nm laser amplitude is controlled using a double-pass AOM with a rise time of $\sim 10~$ns. 
In the ideal model for the quench, the optimal quench time would be $\tau = 4\pi / (3 \Omega \sqrt{3}) = 19~$ns for the high-power Rabi frequency. However, the $10~$ns rise time extends the necessary quench time to the experimentally optimized $\sim 30~$ns. We note that during the rise time, the laser power is increasing to its maximum value, leading to deviations from the ideal model for the quench; this may contribute to a reduction in the measured value of $X-$string  parities.

Throughout this work, measurements of $Z$ and $X$ parities are averaged over identical loops, including reflection and rotation symmetries, across the system. However, loops which touch the edge of the system are excluded to avoid boundary effects.
Error bars are calculated as the standard error of the mean as $\sigma(P)/\sqrt{R}$, where $R$ is the number of repetitions and $\sigma(P)$ is the standard deviation of the parity $P$, which is the average over all identical loops for each repetition. 

\section{Basis rotation for $X$ and $Z$ parity loops}
The basis rotation used to measure $X$ parity loops is applied with a reduced blockade radius which, in the ideal limit, removes interactions between separate triangles while maintaining a hard blockade constraint on Rydberg excitations within single triangles. The rotation can therefore be understood by its action on individual fully-blockaded triangles.
The Hilbert space for each triangle is four-dimensional, allowing for either zero Rydberg excitations, or one Rydberg excitation on any of the three links.
Taking $\includegraphics[valign=c,width=0.13\columnwidth]{triangle_qubits-02.png}$ as the basis states, the Hamiltonian for the quench in the limit of perfect intra-triangle blockade is described by the following matrix:
\begin{equation}
\label{eq:QuenchHamiltonianMatrix}
    H = \frac{\Omega}{2}\begin{pmatrix}
    0 & -i & -i & -i \\
    i & 0 & 0 & 0\\
    i & 0 & 0 & 0\\
    i & 0 & 0 & 0
    \end{pmatrix}
\end{equation}

\begin{figure}[hb!]
\includegraphics[width=3.6in]{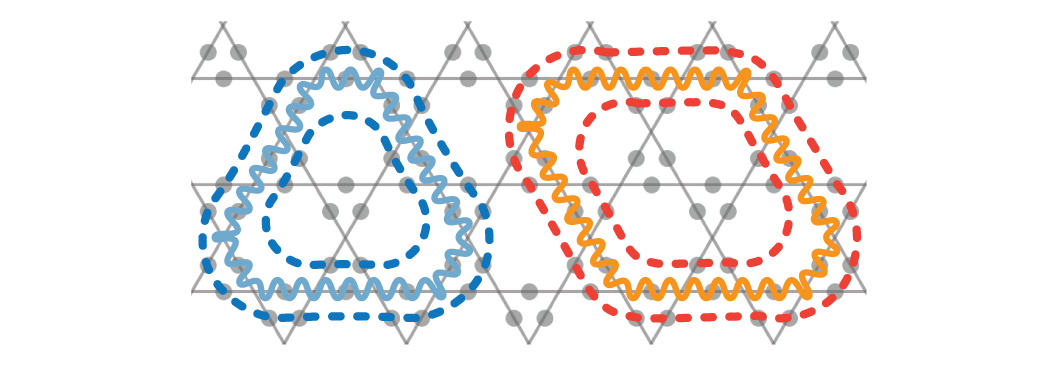}
\caption{\textbf{Dual $Z$ and $X$ loops.} Examples of dual $Z$ loops (dashed lines) to closed $X$ loops (solid wiggly lines).}
\label{fig:DualLoops}
\end{figure}

The basis rotation shown in Fig.~3C of the main text, which relates $X$ and $Z$ parity under evolution through this quench Hamiltonian \eqref{eq:QuenchHamiltonianMatrix}, 
was proven in Ref.~\cite{Ruben2020} by direct computation. Here we provide an alternative derivation. Firstly, we
note that the $Z$ operator acting on the upper two edges of a triangle (\insertimage{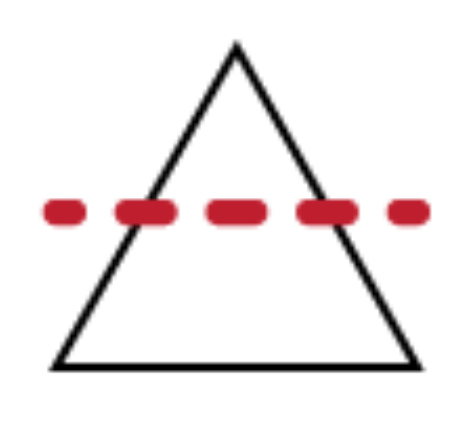}) and the $X$ operator acting on the lower edge of a triangle (\insertimage{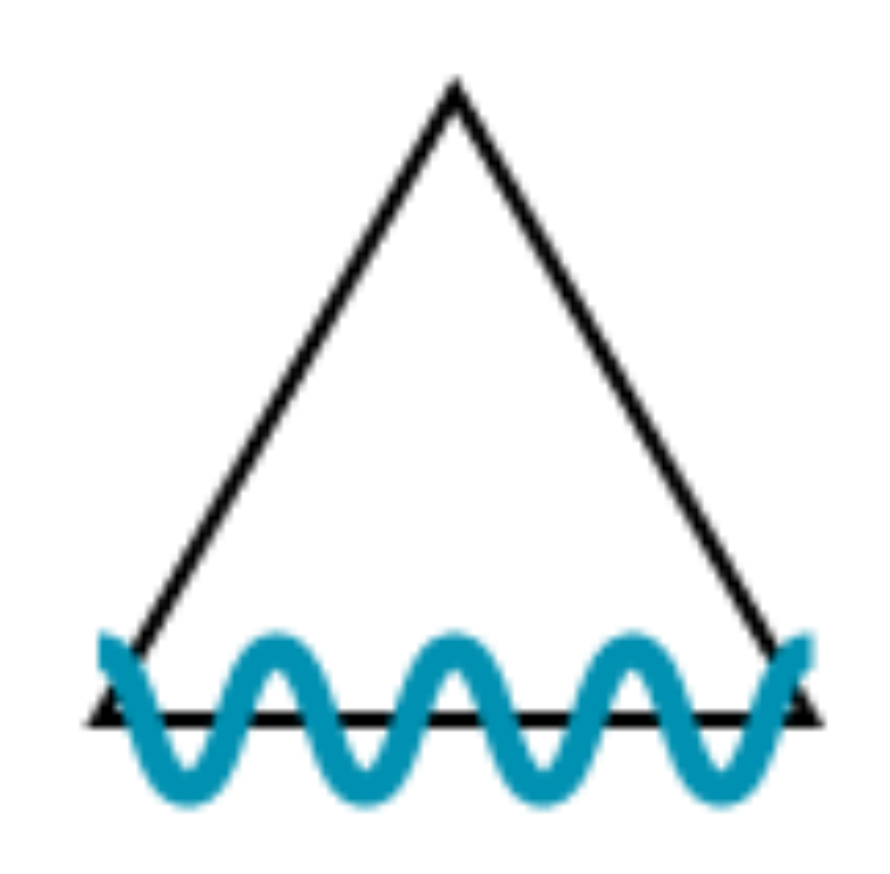}), defined in Figs. 2,3 of the main text, are given by:
\begin{align}
    \insertimage{Z_parity.png} &= \begin{pmatrix}
    1 & 0 & 0 & 0 \\
    0 & 1 & 0 & 0 \\
    0 & 0 & -1 & 0 \\
    0 & 0 & 0 & -1 \\
    \end{pmatrix} \\
    \insertimage{X_parity.png} &= \begin{pmatrix}
    0 & -1 & 0 & 0 \\
    -1 & 0 & 0 & 0 \\
    0 & 0 & 0 & 1 \\
    0 & 0 & 1 & 0 \\
    \end{pmatrix} 
\end{align}
The $X$ and $Z$ parity operators can be mutually diagonalized by changing to an appropriate symmetrized basis:
\begin{center}
\begin{tabular}{c||c|c}
    Basis state & \insertimage{Z_parity.png} & \insertimage{X_parity.png} \\
     \hline
     $\ket{0} = \insertimage{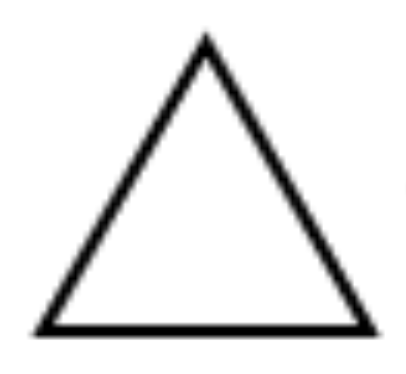} + \insertimage{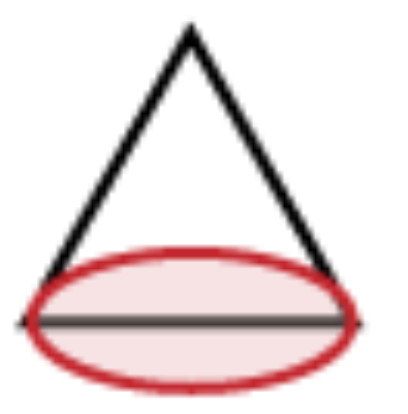}$ & +1 & -1 \\
     $\ket{1} = \insertimage{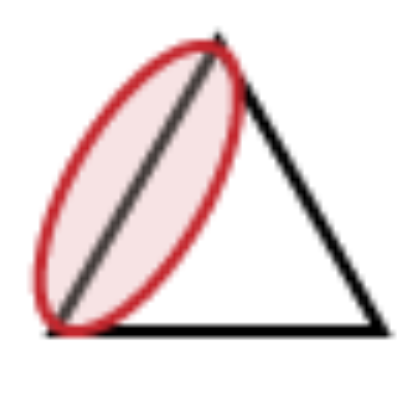} + \insertimage{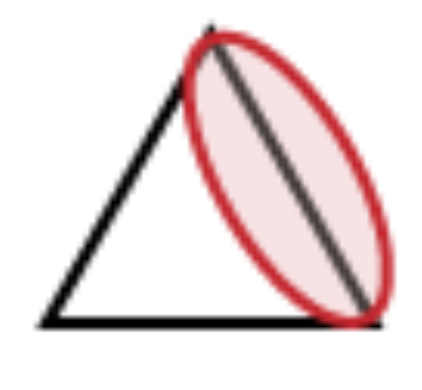}$ & -1 & +1 \\
     $\ket{2} = \insertimage{Triangle_0.png} - \insertimage{Triangle_1.png}$ & +1 & +1 \\
     $\ket{3} = \insertimage{Triangle_2.png} - \insertimage{Triangle_3.png}$ & -1 & -1 \\
\end{tabular}
\end{center}
In this basis, the quench Hamiltonian \eqref{eq:QuenchHamiltonianMatrix} is expressed as:
\begin{equation}
    H = \frac{\Omega}{2} \begin{pmatrix}
    0 & i & -i & 0 \\
    -i & 0 & -i & 0 \\
    i & i & 0 & 0 \\
    0 & 0 & 0 & 0
    \end{pmatrix} \label{eq:rot}
\end{equation}
This Hamiltonian generates cyclic permutations among the basis states $\ket{0}, \ket{1}$, and $\ket{2}$, while leaving $\ket{3}$ invariant. The permutation $\ket{0} \to \ket{1} \to \ket{2} \to \ket{0}$ maps the \insertimage{X_parity.png} eigenvalue to the \insertimage{Z_parity.png} eigenvalue for each initial state.
Moreover, the invariant state $\ket{3}$ has both $\insertimage{X_parity.png} = \insertimage{Z_parity.png} = -1$, so it automatically satisfies the target eigenvalue mapping.
Thus, after an appropriate evolution time corresponding to a single cyclic permutation ($\tau = \frac{4\pi}{3\sqrt{3} \Omega}$), all \insertimage{X_parity.png} eigenvalues have been mapped to \insertimage{Z_parity.png} eigenvalues, which is diagonal in the measurement basis. Formally, this can be expressed as:
\begin{equation}
    \insertimage{X_parity.png} = e^{i H \tau} \insertimage{Z_parity.png} e^{-i H \tau}
\end{equation}
We further note that this relationship holds also for parity operators defined on other sides of the triangle, e.g., $\insertimage{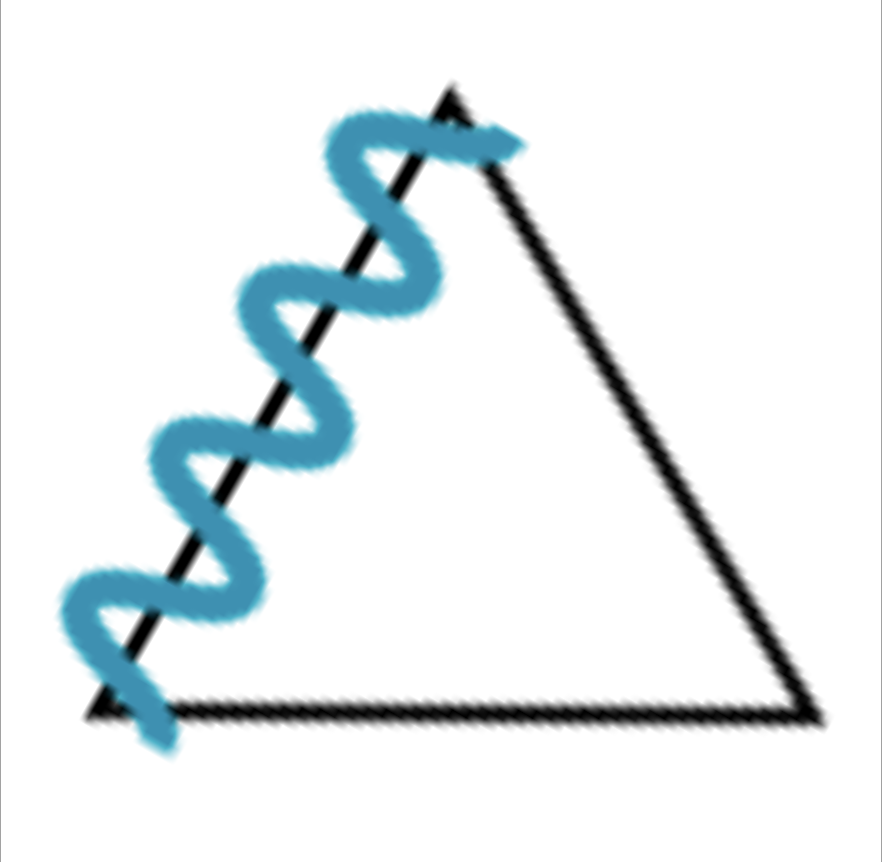} = e^{i H \tau} \insertimage{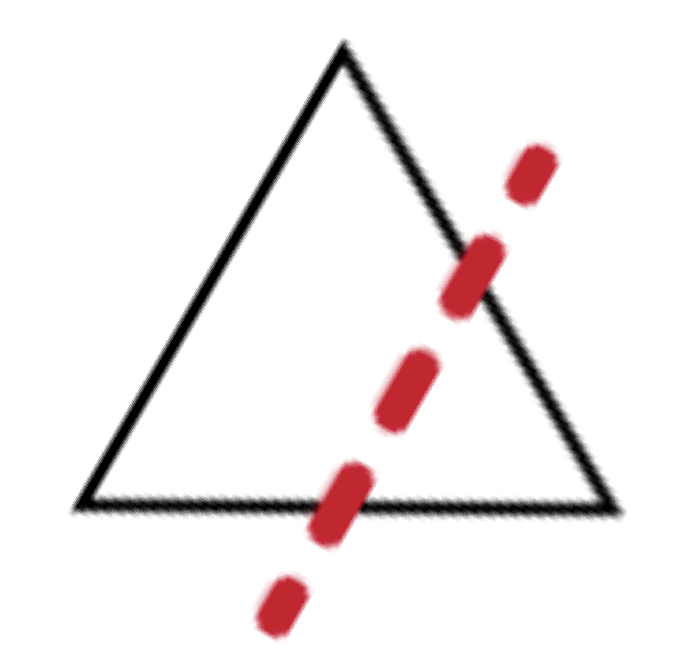} e^{-i H \tau}$. Large $X$ parity strings or loops can be decomposed in terms of their action on individual triangles, and since the basis rotation acts on each triangle individually, this extends the mapping from $X$ strings to corresponding dual $Z$ strings in the rotated basis, as illustrated in Fig.~\ref{fig:DualLoops}.

\section{Supplemental experimental data}

\subsection{Mean Rydberg density and boundary effects}

\begin{figure*}[hb!]
\includegraphics[width=0.7\textwidth]{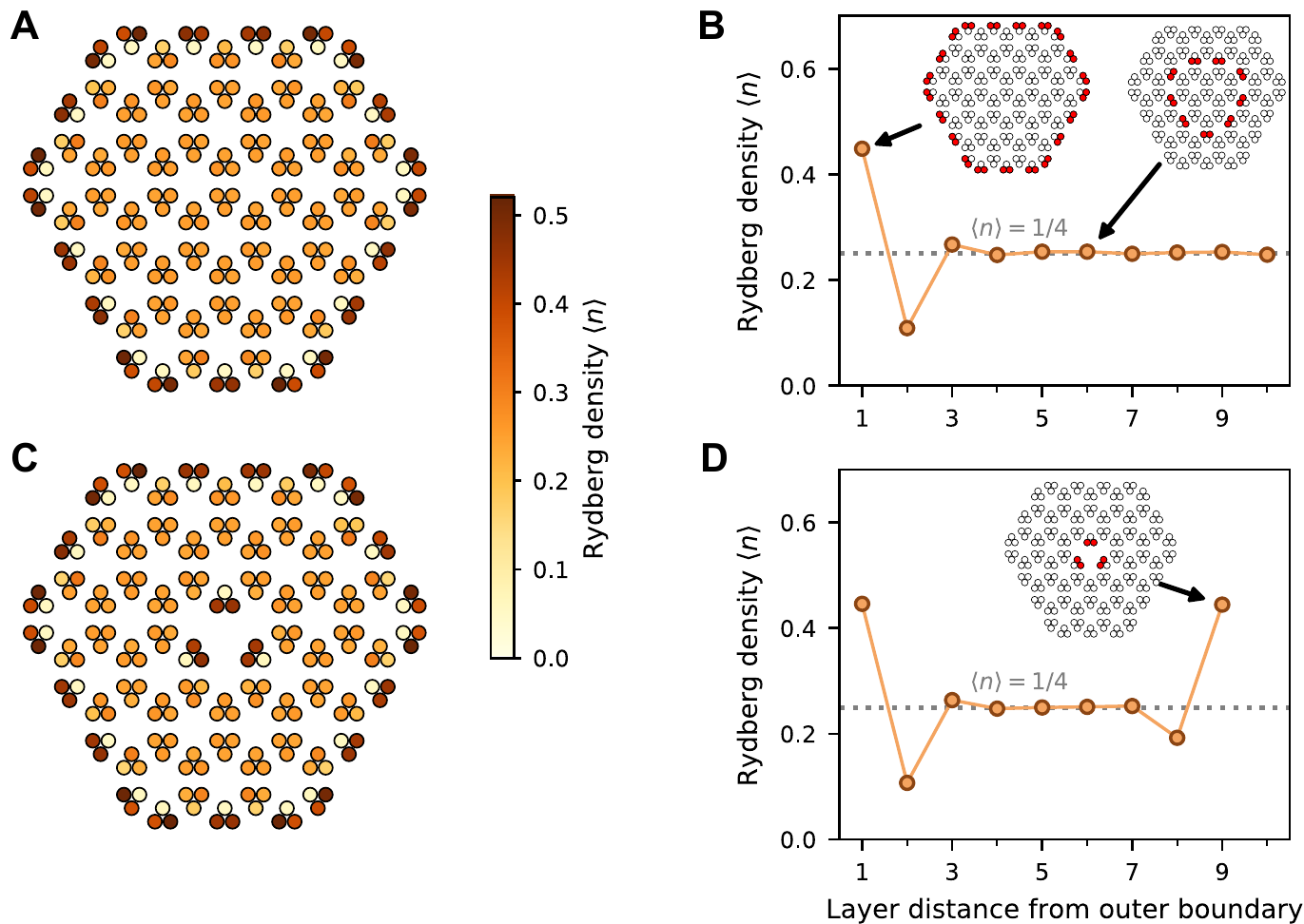}
\caption{\textbf{Site-by-site mean Rydberg density}. We measure the mean Rydberg excitation density $\langle n \rangle$ site-by-site in the dimer phase with $\Delta/\Omega = 4$ for both full arrays (\textbf{A}) as well as arrays with a hole (\textbf{C}). (\textbf{B},\textbf{D}) We then plot the corresponding mean density layer-by-layer as a cross-section from the edge into the bulk, showing that within the outer two to three layers, the bulk settles into the $\langle n \rangle \sim 1/4$ phase.}
\label{fig:MeanRydbergDensities}
\end{figure*}

After preparing the dimer phase for $\Delta/\Omega \sim 4$, we observe a Rydberg excitation density in the bulk of $\langle n \rangle \sim 1/4$. The sites close to the boundary of the system, however, are dominated by edge effects. In Fig.~\ref{fig:MeanRydbergDensities}, we show the Rydberg excitation density site-by-site, and demonstrate that the edge effects only permeate two to three layers into the bulk before the $\langle n \rangle \sim 1/4$ plateau is reached. In arrays with a topological defect, the hole forms an inner boundary and similarly induces edge effects (Fig.~\ref{fig:MeanRydbergDensities}C,D).
These observations allow us to determine the minimum system sizes that may be used such that the physics of the system is not dominated by boundary effects, resulting in our choice of the 219-atom arrays used in this work.

\subsection{Lack of spatial order within spin-liquid phase}

\begin{figure*}[hb!]
\includegraphics[width=0.95\textwidth]{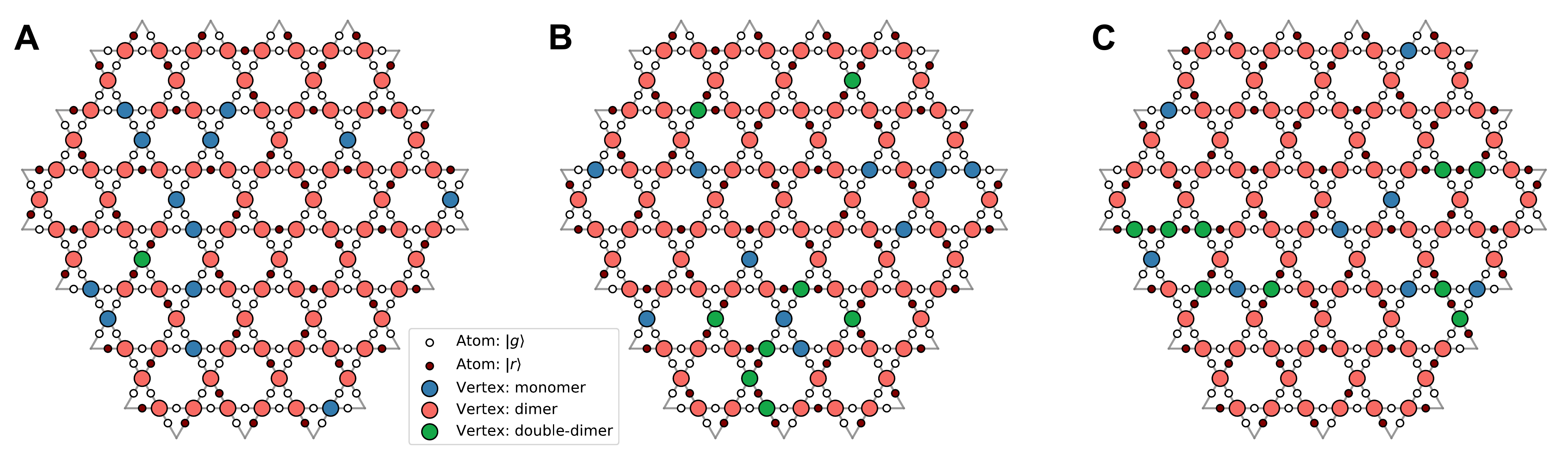}
\caption{\textbf{Snapshots in the dimer phase.} Three sample experimental realizations within the dimer phase at $\Delta/\Omega = 4.3$. The binarized atom readout is shown by small circles on the links of the kagome lattice, with open circles denoting $\ket{g}$ and filled circles denoting $\ket{r}$. Vertices of the kagome lattice (large circles) are colored according to the number of adjacent atoms in $\ket{r}$ to visually accentuate which parts of the system are properly covered with dimers.}
\label{fig:Snapshots}
\end{figure*}

The lack of spatial order in the spin-liquid phase is a key feature that separates this phase from possible nearby solid phases. At the simplest level, spatial order can be assessed by looking at individual projective measurements of the atomic states in the ensemble. We show three examples of such snapshots in Fig.~\ref{fig:Snapshots}, where the measured states of individual atoms are represented as small circles on the links of the kagome lattice, filled or unfilled indicating a Rydberg state or a ground state, respectively. In the mapping to a monomer-dimer model, we can alternatively consider the vertices of the kagome lattice in terms of how many adjacent Rydberg excitations (dimers) are present. In practice, vertices can have zero attached dimers (so-called monomers), a single attached dimer (corresponding to an ideal dimer covering), or more attached dimers (violating the long-range blockade constraint). In Fig.~\ref{fig:Snapshots}, we additionally color each vertex according to the number of such attached dimers. The widespread abundance of vertices connected to a single dimer (Fig.~1E and snapshots from Fig.~\ref{fig:Snapshots}) signifies occupation of the dimer phase.

\begin{figure}
\includegraphics[width=3.0in]{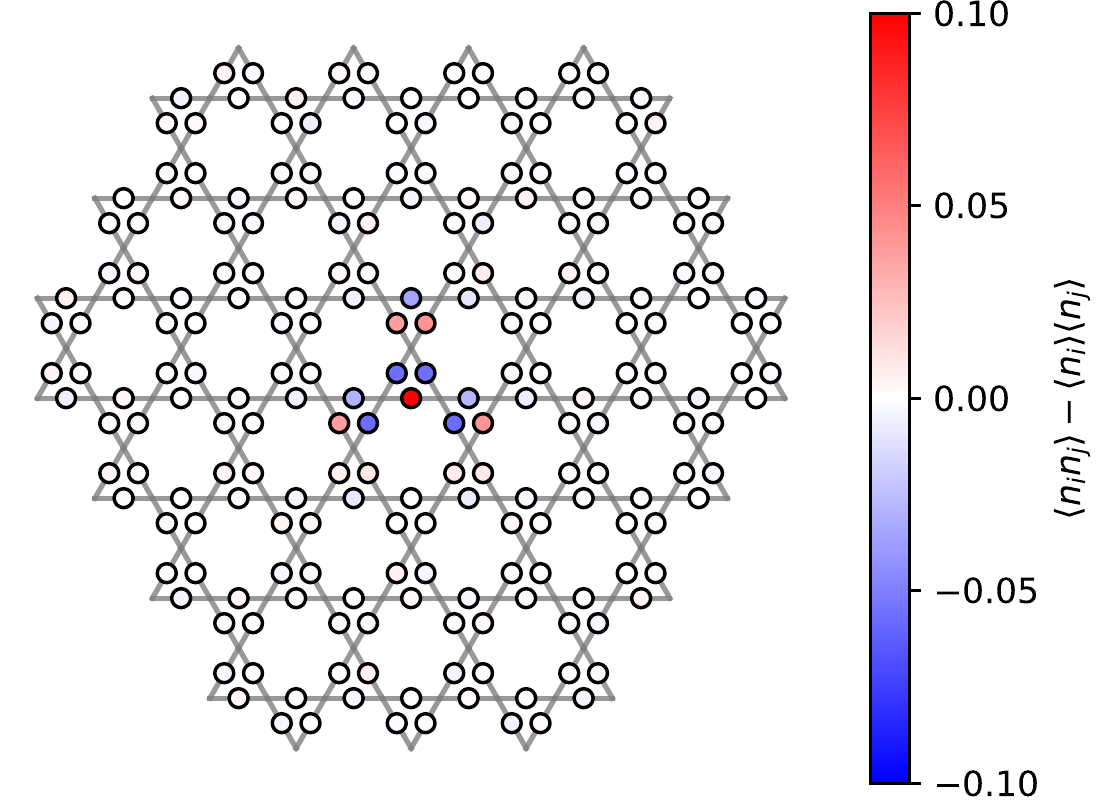}
\caption{\textbf{Density correlations  between individual Rydberg excitations.}
We directly measure the Rydberg density-density correlator $\langle n_i n_j \rangle - \langle n_i \rangle \langle n_j \rangle$ between a central atom and all other atoms in the system. We observe anticorrelations between the central atom and the other two atoms in the same triangle, as well as with atoms in the adjacent triangles, given by the choice of blockade radius $R_b$. Longer range correlations vanish. This data was taken at $\Delta / \Omega = 4.3$.}
\label{fig:Correlations}
\end{figure}

Moreover, spatial correlations can be used to look for solid-type spatial order (Fig.~\ref{fig:Correlations}). We measure 
Rydberg density-density correlations on the atomic array and find non-vanishing correlations for atoms within a single triangle or between adjacent triangles, with vanishing correlations over longer distances. This observation confirms the lack of spatial order in the dimer phase we prepare.

\subsection{Phase dependence of quench}

\begin{figure}[hb!]
\includegraphics[width=\columnwidth]{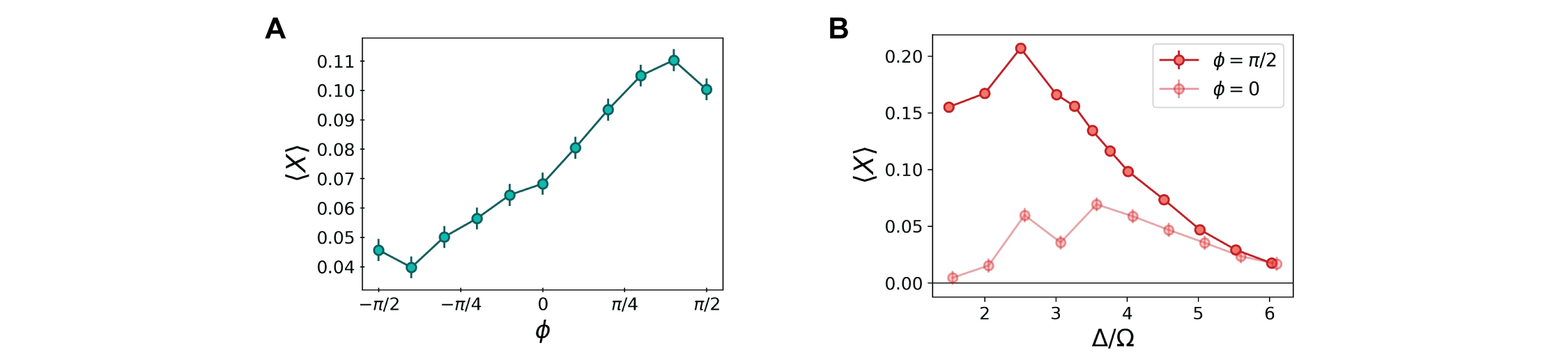}
\caption{\textbf{Phase dependence of quench.} (\textbf{A}) After preparing the dimer phase at $\Delta / \Omega = 4$, we quench for the pre-calibrated time $\tau$ with a variable quench phase and measure the resulting $X$ loop parity around a single hexagon. (\textbf{B}) For fixed quench phase $\phi=\pi/2$ or $\phi=0$, we measure the $X$ parity after the pre-calibrated quench time as a function of the final detuning of the cubic sweep. The data for $\phi=\pi/2$ is reproduced from Fig.~3F of the main text.}
\label{fig:PhaseDependence}
\end{figure}

The quench which induces the basis rotation for measuring $X$ parity is implemented by rapidly switching the laser detuning to $\Delta_q = 0$ following the preparation of the dimer phase, and simultaneously changing the phase of the laser field by $\pi/2$. This choice of phase approximately maximizes the $X$ parity signal, as measured by applying the same quench duration but with variable phase (Fig.~\ref{fig:PhaseDependence}A). 

The phase change can be understood by interpreting it as evolution under $\sum_i n_i$ for time $\phi$, followed by a fixed-phase quench. Since the quench ultimately measures coherences between different components of the wavefunction, this phase change only matters insofar as it changes the relative phases between components. We note here that coherences between perfect dimer coverings will be unchanged by the phase change, since all perfect dimer coverings have the same number of Rydberg excitations. A wavefunction which is the superposition of all perfect dimer coverings, then, would be insensitive to the choice of phase for the quench. However, in our system there is a finite density of both monomers and vertices with two attached dimers. An $X$ loop crossing through a monomer creates a double-dimer at that vertex, and these types of  component pairs are additionally included in our $X$ parity measurements. Since the coupled states with a monomer and a double-dimer have different numbers of Rydberg excitations, these coherences \emph{are} phase-sensitive. Comparing the measured $X$ parity for $\phi=\pi/2$ and $\phi=0$ as we scan across the phase diagram (Fig.~\ref{fig:PhaseDependence}B), we find that the first has larger amplitude and extends more strongly into the trivial phase, consistent with the expectation from theoretical calculations \cite{Ruben2020}.

\subsection{$Z$ parity measurements with improved state preparation}

\begin{figure}[hb!]
\includegraphics[width=3.6in]{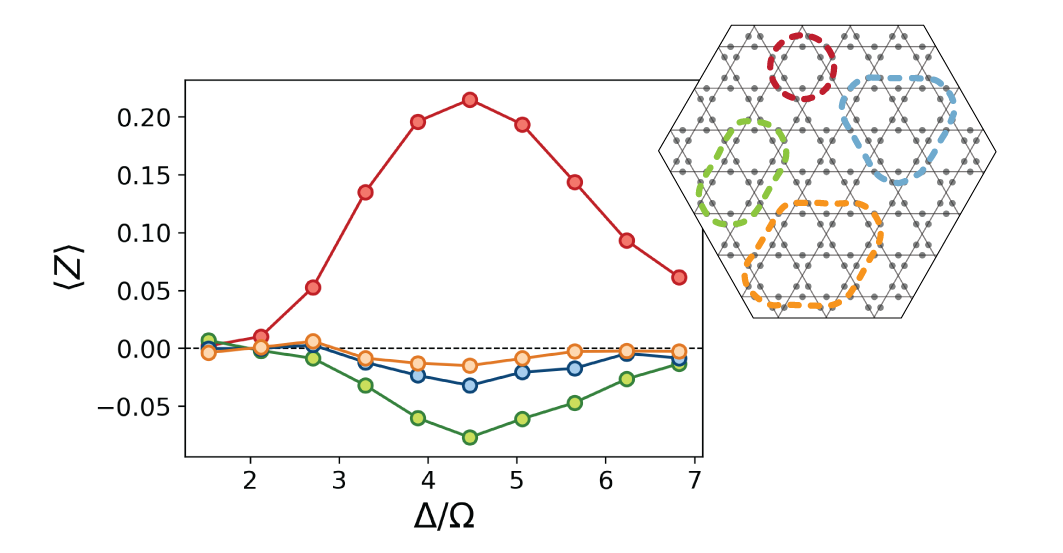}
\caption{\textbf{$Z$ loop parity with improved state preparation.} We measure $Z$ on closed loops with a larger intermediate state detuning for the two-photon Rydberg excitation to reduce spontaneous emission rates, and with a larger Rabi frequency during the state preparation. We observe larger parities than in the comparable Fig.~2 of the main text. }
\label{fig:Z_1GHz}
\end{figure}

All data shown in the main text is taken with intermediate detuning $\delta = 2\pi \times -450~$MHz (see Sec.~\ref{Sec:expt}) for the two-photon Rydberg excitation. This choice is to enable our largest dynamic range of Rabi frequencies, which is crucial for being able to perform state preparation at low $\Omega$ and then apply the quench at large $\Omega$ with reduced blockade radius. Larger intermediate detuning would require performing state preparation at an even lower initial Rabi frequency, where we observe worse results.
However, the small intermediate detuning introduces stronger decoherence due to increased spontaneous emission from the intermediate state. To supplement these results, we additionally perform state-preparation and measure $Z$ parity at an increased intermediate detuning of $\delta = 2\pi \times 1$~GHz. To further optimize this state preparation, we use a larger Rabi frequency $\Omega=2\pi\times1.7$ MHz and a smaller lattice spacing $a=3.7~\mu$m, which should improve adiabaticity during the preparation.  In this configuration, we indeed observe larger $Z$ loop parities (Fig.~\ref{fig:Z_1GHz}), but we cannot measure corresponding $X$ loop parities. This highlights that the large dynamical range required for the measurement of the $X$ operator is one of the main technical challenges of this experimental work. At the same time, it shows that with more available laser power for Rydberg excitation, the quality of state preparation can be improved by working at this increased intermediate detuning and higher Rabi frequencies (and with smaller lattice spacings to achieve the same blockade radius). 

\subsection{Correlations between parity loops}
String operators are used in this work to assess long-range topological order. However, the large loops which are studied can be decomposed into the product of smaller loops around sub-regions: for example, $X$ loops can be decomposed into the product of enclosed hexagons. To demonstrate that the parity measured on large loops is indeed indicative of long-range order, rather than emerging from the ordering of each hexagon individually, we extract correlations between the separate parity loops which comprise larger loops. 

We first study parity loops which enclose adjacent hexagons of the kagome lattice. The minimal such $X$ parity loop is exactly equal to the product of the parity around the two enclosed hexagons. The connected correlator of the parity around these two inner hexagons is
\begin{equation}
    G^{(2)}_X = \langle \insertimage{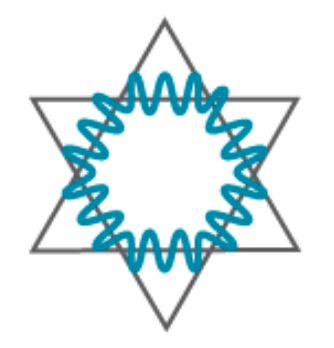}_1\insertimage{X_around_hex.png}_2 \rangle
    - \langle \insertimage{X_around_hex.png}_1 \rangle \langle \insertimage{X_around_hex.png}_2 \rangle
\end{equation}
Similarly, $Z$ loops which enclose two hexagons decompose into the product of $Z$ parity around the two hexagons, multiplied additionally by the parity around the central interior vertex (which should always be -1 in a dimer covering). We define the analogous two-hexagon connected correlator for $Z$ as
\begin{equation}
    G^{(2)}_Z = \langle \insertimage{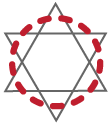}_1 \insertimage{single_star}_2 \rangle - \langle \insertimage{single_star}_1 \rangle \langle \insertimage{single_star}_2 \rangle
\end{equation}

Higher-order connected correlations between three adjacent hexagons which form a triangle further highlight nonlocal correlations in this system.
We define the connected three-point correlator \cite{rispoli_quantum_2019} which subtracts away contributions from underlying two-point correlations as 
\begin{align}
\begin{split}
    G^{(3)}_X &= \langle \insertimage{X_around_hex}_1 \insertimage{X_around_hex}_2 \insertimage{X_around_hex}_3 \rangle  - G^{(2)}_{X,12} \langle \insertimage{X_around_hex}_3 \rangle - G^{(2)}_{X,23} \langle \insertimage{X_around_hex}_1 \rangle - G^{(2)}_{X,31} \langle \insertimage{X_around_hex}_2 \rangle - \langle \insertimage{X_around_hex}_1 \rangle \langle \insertimage{X_around_hex}_2 \rangle \langle \insertimage{X_around_hex}_3 \rangle
\end{split}
\end{align}
where $G^{(2)}_{X,ij}$ is the connected correlator for hexagons $i,j$. Third order connected correlators for $Z$ parity are analogously defined.

\begin{figure}[hb!]
\includegraphics[width=\columnwidth]{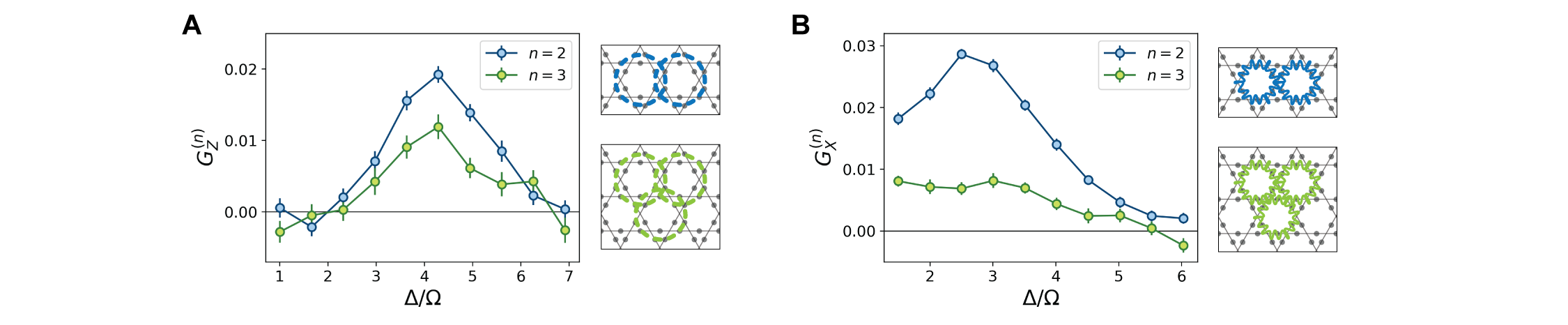}
\caption{\textbf{Correlations between parity loops.} We measure two-point and three-point connected correlations between parity around adjacent hexagons. (\textbf{A}) $Z$ parity correlations between loops which enclose pairs and triplets of adjacent hexagons. (\textbf{B}) $X$ parity correlations between pairs and triplets of adjacent hexagons.}
\label{fig:Loops_correlators}
\end{figure}

As shown in Fig.~\ref{fig:Loops_correlators}, we observe nonzero two-hexagon and three-hexagon connected correlations within the dimer phase region, indicating that the parity measured on double-hexagon and triple-hexagon loops does not emerge from independently determined parity around each interior subregion, but instead emerges due to nontrivial  correlations over longer length scales.

\subsection{Quasiparticle excitations}

Within the dimer-monomer model, quasiparticle excitations of two types are created by the application of open $X$ and $Z$ strings: these are the electric ($e$) and magnetic ($m$) anyons, respectively. Open $X$ strings create monomers (or double-dimers) at their endpoints, and thus $e$-anyons are identified as defects in the dimer covering. 
\begin{figure}[hb!]
\includegraphics[width=\columnwidth]{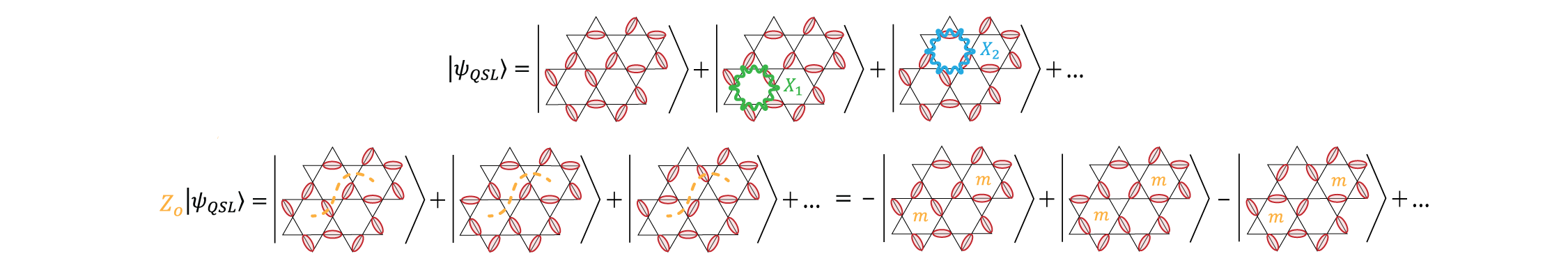}
\caption{\textbf{Magnetic anyons.} The dimer states contained in $|\psi_{QSL}\rangle$ are connected to each other by the application of $X$ on closed loops. When an open $Z$ string acts on the superposition, the dimer states connected by an open loop that encloses one end of the string ($X_1$) acquire opposite signs. The $m$-anyons generated at the endpoints of the string are then detected by $X$ loops that enclose one of them, e.g. $\langle X_1 \rangle=-1$, while $\langle X_2 \rangle=+1$ is unperturbed.}
\label{fig:m_anyons}
\end{figure}
Open $Z$ strings on the other hand impart a relative phase between various dimer configurations, corresponding to $m$-anyons. To understand $m$-anyons, we first note that all dimer coverings in the QSL superposition are related to one another by the application of properly chosen closed $X$ loops (first row of Fig.~\ref{fig:m_anyons}). An open $Z$ string applied to the QSL state, then, results in different dimer coverings acquiring $\pm$ phase factors according to the number of dimers crossed by the string. Whenever two dimer configurations are related by a closed $X$ loop which encloses one of the endpoints of the $Z$ string, they acquire opposite signs (Fig.~\ref{fig:m_anyons}). After the application of the open $Z$ string, then, $\langle X \rangle$ is inverted for any closed loop around one endpoint of the $Z$ string, analogous to how  $\langle Z \rangle$ around the endpoint of an open $X$ string (a defect) is inverted. Since open $Z$ strings terminate in the hexagons of the kagome lattice, we associate the resulting magnetic ($m$) anyons as living on these hexagons, and the $X$ parity around hexagons therefore detects the presence of $m$-anyon excitations.

\begin{figure}[h!]
\includegraphics[width=\columnwidth]{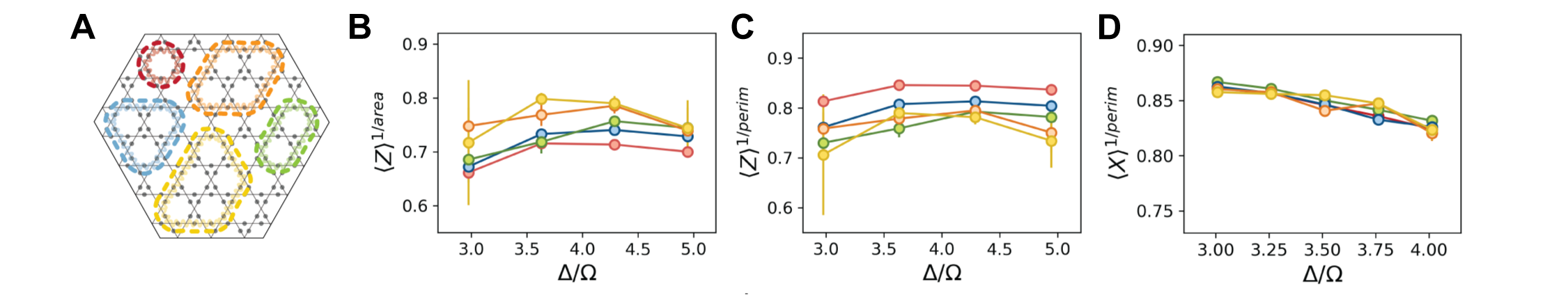}
\caption{\textbf{Scaling of $Z$ and $X$ parities with the loop size.} We calculate the rescaled parities $\langle Z \rangle^{1/\textnormal{area}}$ (\textbf{B}), $\langle Z\rangle^{1/\textnormal{perim}} $ (\textbf{C}) and $\langle X \rangle^{1/\textnormal{perim}}$ (\textbf{D}) for the different loop sizes in (\textbf{A}). While for the $X$ operator we observe a very good perimeter law scaling on the entire range of detunings, for $Z$ we observe an approximate area law scaling for small loop sizes that finally converges towards a perimeter law scaling.
}
\label{fig:Loops_scaling}
\end{figure}

In Fig.~\ref{fig:Loops_scaling} we report the $Z$ and $X$ loop parities rescaled with area and perimeter law for different values of $\Delta$ in the relevant range of detunings. We observe that the excellent perimeter law scaling of $X$ reported in Fig.~4I of the main text extends over the entire range of $\Delta$. For $Z$ instead we find that the initial approximate area law scaling converges towards a perimeter law for large loops.

We can shed light on the scaling behavior observed in the experiment by comparing it with the expected scaling from theory. Let us first note that the generic equilibrium expectation for both string operators is a perimeter law scaling \cite{Gregor11}. This can be seen as a consequence of the mutual statistics of $e$- and $m$- anyons: since there will be virtual fluctuations of both anyons, these will induce correlations\footnote{To clarify this further, we note that the monomers (and double-dimers) visible in the experimental snapshots need not to directly correspond to physical excitations, since the ground state will have so-called `virtual' fluctuations when it is not in an idealized fixed-point state. These can be interpreted  as correlated $e$-anyons. In contrast, in an ideal $\mathbb Z_2$ spin liquid, \emph{physical} $e$-anyons will be uncorrelated---since this is a defining property of the deconfined phase where $e$-anyons move independently at sufficiently large distances.}
for anyons of the other type, leading to a perimeter law. This generic expectation of a perimeter law is well-known in the (lattice) gauge theory community, and can be related to the phenomenon of string breaking \cite{Fradkin79}. Experimentally, we observe a perimeter law for $X$-loops and an (approximate) area law for $Z$-loops (with substantial deviations for larger loop sizes). This  can be understood by noting that we enter the QSL-like state from the trivial phase, which can be interpreted as a condensate of $e$-anyons (i.e., both closed and open $X$-strings give nonzero correlations): the perimeter law for closed  $X$ strings is thus already present in the trivial phase and naturally persists into the QSL-like state (while correlations for the \emph{open} $X$-strings vanish). In contrast, the $Z$-correlations are absent in the trivial phase proximate to the QSL: these are only developed at the quantum critical point, and since we sweep through this at a finite rate, the $Z$-loop correlations are only developed over a characteristic length scale, implying an area law. Numerically, we indeed confirm that $Z$-loop correlations are significantly enhanced upon increasing preparation time (see Sec.~\ref{Theory}), consistent with our observations in Fig.~\ref{fig:Z_1GHz}. We note that this imperfect generation of $Z$-loop correlations can be equivalently interpreted as generating a density of $e$-anyon excitations. Dynamically inducing the onset of a QSL and possible meta-stable states are rich phenomena which deserve further detailed study.

\subsection{Additional data for arrays with nontrivial topology}

\begin{figure}[h!]
\includegraphics[width=3.6in]{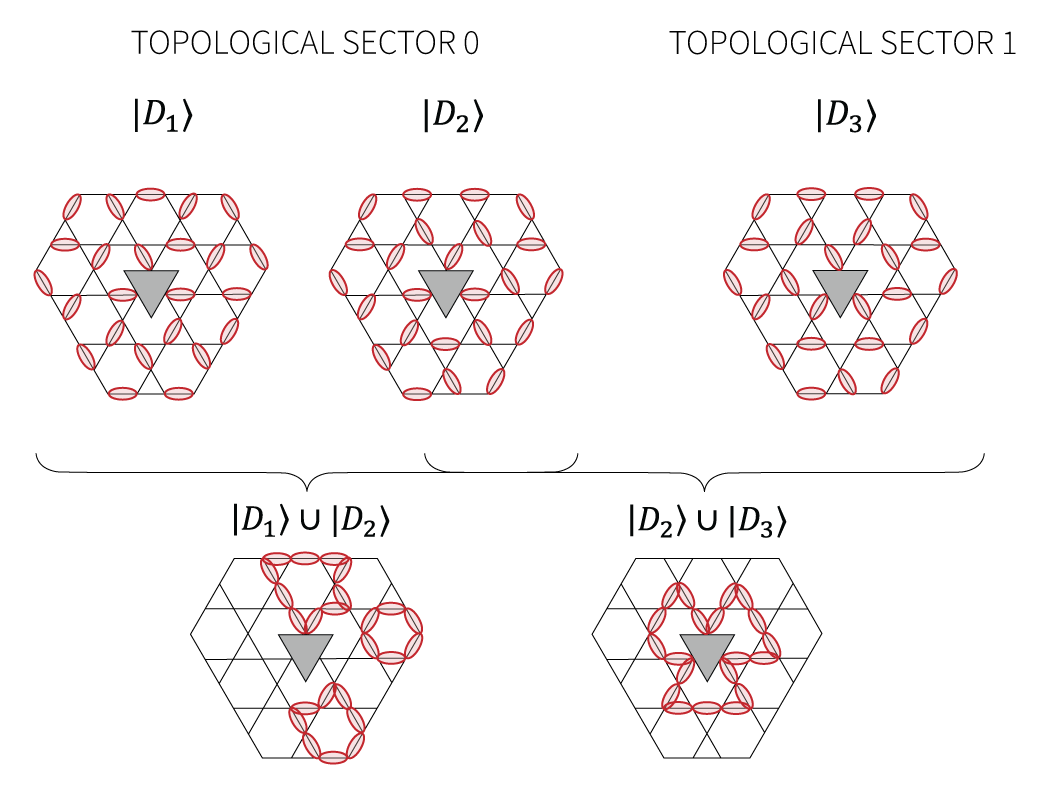}
\caption{\textbf{Distinction between topological sectors.} To determine if the three dimer coverings $|D_1\rangle$,$|D_2\rangle$ and $|D_3\rangle$ belong to the same or opposite topological sectors, we build the transitions graphs $|D_1\rangle \cup |D_2\rangle$ and $|D_2\rangle \cup |D_3\rangle$. In the latter we see that the dimers form a closed loop around the hole, highlighting that the two states belong to opposite sectors.}
\label{fig:TopSectors}
\end{figure}

The distinction between two distinct topological sectors can be better understood by looking at the transition graphs between pairs of dimer states \cite{Sutherland88}. These are built by superimposing two dimer coverings and removing the overlapping dimers (Fig.~\ref{fig:TopSectors}). The dimer states belong to opposite topological sectors if the remaining dimers form an odd number of closed loops around the hole, indicating the set of non-local moves required to transform one into the other.

\begin{figure}[h!]
\includegraphics[width=3.6in]{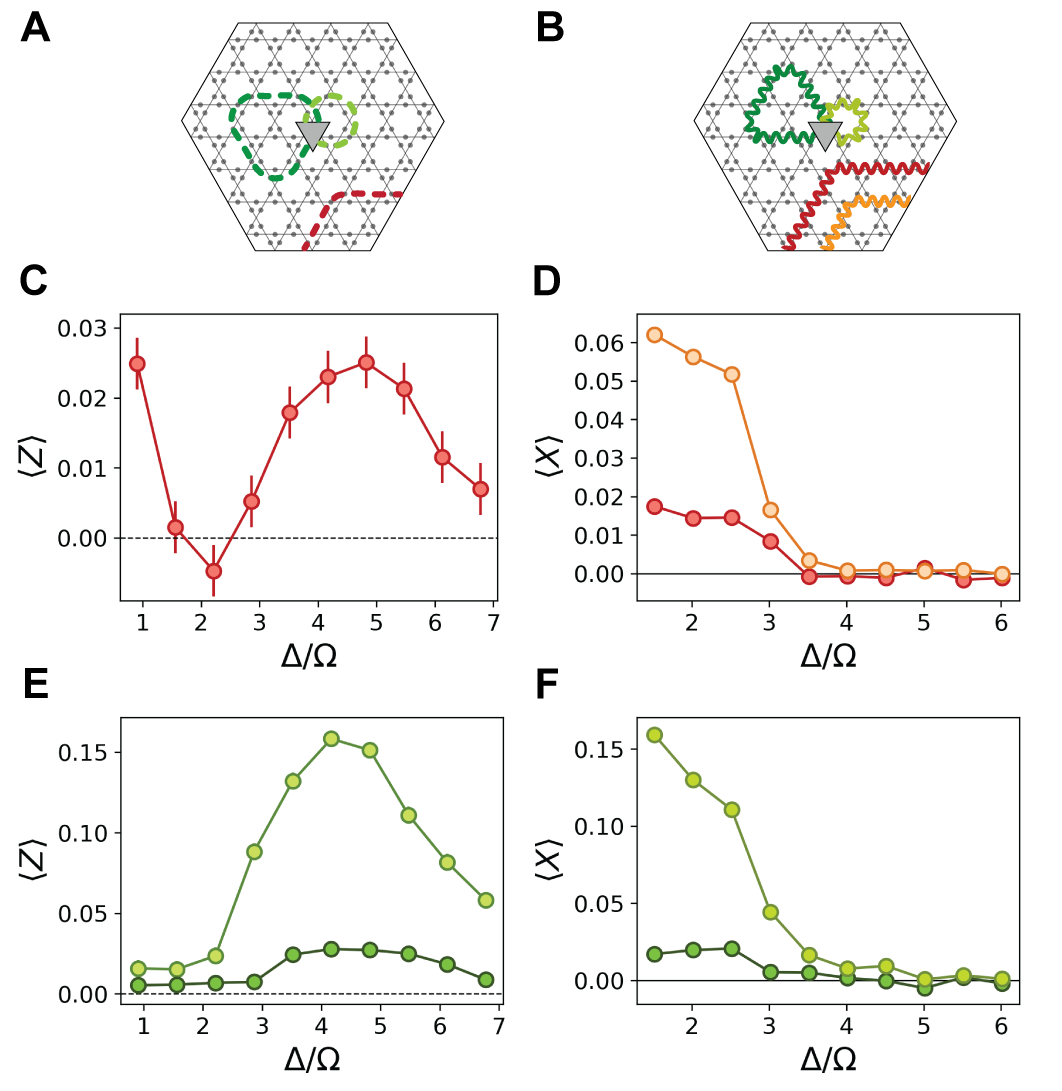}
\caption{\textbf{Boundary-to-boundary string operators.} We measure the $Z$ (\textbf{A}) and $X$ (\textbf{B}) operators on open strings connecting two points on the outer (\textbf{C},\textbf{D}) or inner (\textbf{E},\textbf{F}) boundaries of the array. Observing the same features for both, we confirm that the small central hole does indeed create an effective inner boundary.}
\label{fig:SuppHole}
\end{figure}

To demonstrate that the removal of three atoms at the center of the array creates an actual inner boundary, we measure the $Z$ and $X$ operators on strings with both endpoints on the inner or outer boundaries (Fig.~\ref{fig:SuppHole}). In the relevant range of detunings ($3.3 \lesssim \Delta/\Omega \lesssim 4.5$) we measure a finite $\langle Z\rangle$ and a vanishing $\langle X \rangle$ in both cases, indicating that the central hole also generates an effective boundary. This also confirms that the boundaries that are naturally created in our system are of the $m$-type, i.e. $m$-anyons localize on it (hence the finite $\langle Z\rangle$) \cite{Ruben2020}. 

\section{Numerical Studies \label{Theory}} 
Below, we report on numerical studies of the Rydberg atom array. We first discuss the zero temperature equilibrium phase diagram, established using density-matrix-renormalization-group (DMRG). Next, we directly simulate the quasi-adiabatic sweep, using both exact diagonalization and dynamical DMRG calculations. To minimize boundary effects due to limitations of numerically accessible system sizes, these calculations are performed on a torus (exact diagonalization) or on an infinite cylinder (DMRG).

\subsection{Ground state phase diagram \label{sec:gs}}

To a first approximation, the Hamiltonian in the main text can be described by an effective `PXP' model \cite{Turner18} 
\begin{equation}
    H_\textrm{PXP} =  \sum_i \left( \frac{\Omega}{2} P \sigma^x_i P - \Delta n_i \right). \label{eq:PXP}
\end{equation}
Here, $P$ is a projector onto $\ket{g}$ for all sites within the blockade radius $R_b$ of the site $i$. This model approximates the the Rydberg Hamiltonian by treating all pairwise interaction energies as either infinite, if within the blockade radius, or zero if beyond. For $R_b = 2.4a$ (as in the main text), this corresponds to blockading the first three interaction distances.
In Ref.~\cite{Ruben2020}, it was shown that this `blockade model' hosts a $\mathbb Z_2$ spin liquid for $1.5 \lesssim \Delta/\Omega \lesssim 2$. 

To include the full van der Waals interactions, we incorporate $V(r) = \Omega(R_b/r)^6$ in the microscopic model within a truncation distance $R_\textrm{trunc}$ (beyond which $V(r)=0$), with $R_b = 2.4a$. On a technical note, we replace the very strong nearest-neighbor repulsion $V(a)/\Omega = (R_b/a)^6 \approx 191$ by $V(a) = +\infty$ by working in an effectively constrained model where any triangle can host at most one dimer. The DMRG \cite{White92,White93} simulations on cylinder geometries \cite{Stoudenmire12} were performed using the Tensor Network Python (TeNPy) package developed by Johannes Hauschild and Frank Pollmann \cite{Hauschild18}. A bond dimension $\chi=1000$ was sufficient to guarantee convergence for the systems and parameters considered.

\begin{figure*}[hb!]
    \centering
    \begin{tikzpicture}
    \node at (0,0) {\includegraphics[scale=0.35]{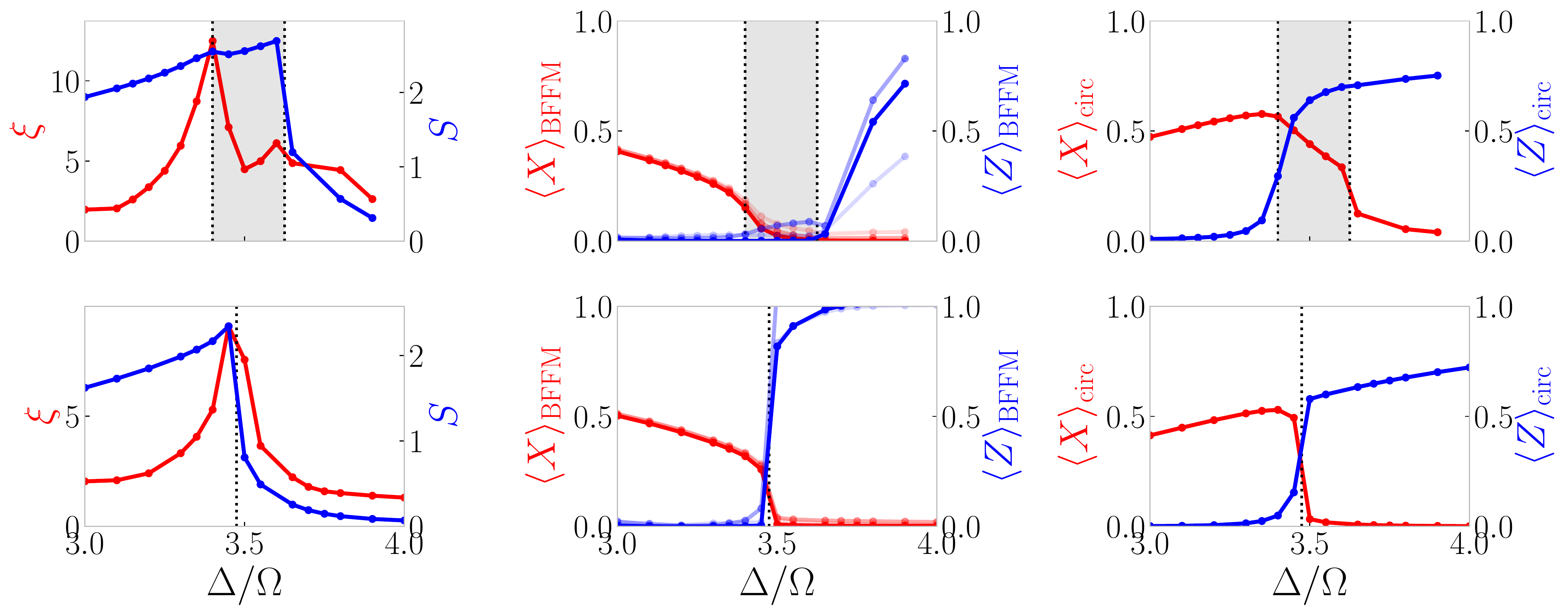}};
    \node at (-7.2,2.7) {\textbf{A}};
    \node at (-7.3+4.8,2.7) {\textbf{B}};
    \node at (-7.3+4.9*2,2.7) {\textbf{C}};
    \node at (-7.2,0.1) {\textbf{D}};
    \node at (-7.2+4.7,0.1) {\textbf{E}};
    \node at (-7.2+9.7,0.1) {\textbf{F}};
    \end{tikzpicture}
    \caption{\textbf{Ground state phase diagram of the link-kagome model for two truncation distances.} All data is for the van der Waals model with blockade radius $R_b =2.4a$ on an XC-8 cylinder. (\textbf{A}--\textbf{C}) For truncation distance $R_\textrm{trunc}=\sqrt{7}a$, we observe a spin liquid (gray shaded area) in between the trivial phase and valence bond solid (VBS). In particular, it is characterized by a large entanglement plateau ($S$ is the entanglement entropy upon bipartitioning the cylinder and $\xi$ is the correlation length), vanishing of the BFFM string order parameters (darker lines correspond to larger strings) and nonzero loop variables ($\langle Z\rangle_{\textrm{circ}}$ and $\langle X\rangle_{\textrm{circ}}$) around the circumference---the signs of the latter label topologically degenerate ground states, as explained in Ref.~\cite{Ruben2020}. (\textbf{D}--\textbf{F}) By increasing the truncation distance to $R_\textrm{trunc}=6a$, the intermediate spin liquid has vanished: there is now a direct first order transition from the trivial phase to a VBS.}
    \label{fig:phasediagram}
\end{figure*}

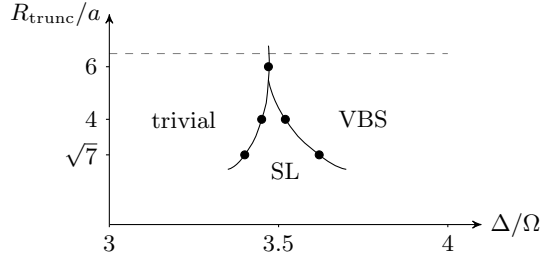
\begin{figure}
    \centering
    \begin{tikzpicture}[xscale=4.5,yscale=0.35]
    \draw[->] (3,0) -- (4.1,0) node[right] {$\Delta/\Omega$};
    \draw[->] (3,0) -- (3,8) node[left] {$R_\textrm{trunc}/a$};
    \draw[-,style=dashed,opacity=0.5] (3,6.5) -- (4,6.5);
    \draw plot [smooth,tension=1] coordinates {(3.35,2.1) (3.4,2.65) (3.45,4) (3.47,5.5) (3.47,6.8)};
    \draw plot [smooth,tension=1] coordinates {(3.7,2.1) (3.62,2.65) (3.52,4) (3.47,5.5)};
    \node[draw, circle, fill=black, minimum size=3pt,style={inner sep=0pt}] at  (3.47,6) {};
    \node[draw, circle, fill=black,minimum size=3pt,style={inner sep=0pt}] at  (3.45,4) {};
    \node[draw, circle, fill=black,minimum size=3pt,style={inner sep=0pt}] at  (3.52,4) {};
    \node[draw, circle, fill=black,minimum size=3pt,style={inner sep=0pt}] at  (3.4,2.65) {};
    \node[draw, circle, fill=black,minimum size=3pt,style={inner sep=0pt}] at  (3.62,2.65) {};
    \draw[-] (3,0) -- (3,-0.1) node[below] {$3$};
    \draw[-] (3.5,0) -- (3.5,-0.1) node[below] {$3.5$};
    \draw[-] (4,0) -- (4,-0.1) node[below] {$4$};
    \draw[-] (3,2.65) -- (3-0.01,2.65) node[left] {$\sqrt{7}$};
    \draw[-] (3,4) -- (3-0.01,4) node[left] {$4$};
    \draw[-] (3,6) -- (3-0.01,6) node[left] {$6$};
    \node at (3.22,4) {trivial};
    \node at (3.75,4) {VBS};
    \node at (3.52,2.05) {SL};
    \end{tikzpicture}
    \caption{\textbf{Ground state phase diagram of the link-kagome model.} Upon including all $V(r) \sim 1/r^6$ interactions (represented by the gray dashed line), we find that there is a direct phase transition from the trivial disordered phase to a crystalline-symmetry-breaking valence bond solid (VBS). However, the model is very close to a spin liquid phase: in fact, if we truncate the interactions to a distance $R_\textrm{trunc}$, we see that a $\mathbb Z_2$ spin liquid can arise in the ground state phase diagram (black dots denote phase transitions obtained via DMRG on the XC-8 cylinder). It is conceivable that dynamical state preparation is not sensitive to the longer-range couplings which destabilize the spin liquid; indeed, in Fig.~\ref{fig:state_prep_DMRG} we confirm that finite-time state preparation gives a state with properties characteristic of a spin liquid. 
    }
    \label{fig:gsphasediagram}
\end{figure}

For intermediate truncation distances, we find a spin liquid in the ground state phase diagram. In particular, taking $R_\textrm{trunc} = \sqrt{7}a\approx 2.65a$, we include four nearest neighbor interactions (i.e., one more than the blockade model): every site is coupled to 10 other sites. The resulting phase diagram is shown in Fig.~\ref{fig:phasediagram}(A--C). This is obtained using the DMRG method applied to an infinitely-long cylinder XC-8 (see Ref.~\cite{Ruben2020} for details about cylinder geometries of the kagome lattice). The presence of a spin liquid is determined based on the behavior of the string observables, as in the experiment. Moreover, we observe topologically degenerate ground states on the cylinder \cite{Ruben2020}.

However, we find that the spin liquid is destabilized upon including even longer range interactions: for $R_\textrm{trunc} =\sqrt{7}a$ we find a spin liquid for $3.4\lesssim \Delta/\Omega \lesssim 3.62$, for $R_\textrm{trunc} = 4a$ we find that this has shrunken down to $3.45\lesssim \Delta/\Omega \lesssim 3.52$, and for $R_\textrm{trunc} = 6a$ there is no intervening spin liquid. Fig.~\ref{fig:phasediagram}(D--F) shows a direct first order phase transition at $\Delta/\Omega \approx 3.47$ from the trivial phase to a valence bond solid (VBS). These results are summarized in Fig.~\ref{fig:gsphasediagram}. We note that these conclusions are strictly valid for the Hamiltonian in eq.~(1) of the main text and might be affected by additional terms, associated, e.g., with multi-body Rydberg interactions. Moreover, other modified ruby lattice geometries still support a ground state spin liquid phase even in the presence of these long range interactions \cite{Ruben2020}. At the same time,  we find that quasi-adiabatic state preparation used in the experiment is far more robust to these effects. In particular, as we will now show, such  state preparation avoids the first order transition to the VBS and instead results in a state reflecting correlations characteristic of a quantum spin liquid.

\subsection{Numerical simulations of dynamical state preparation}

\begin{figure*}[hb!]
    \centering
    \includegraphics[width=15cm]{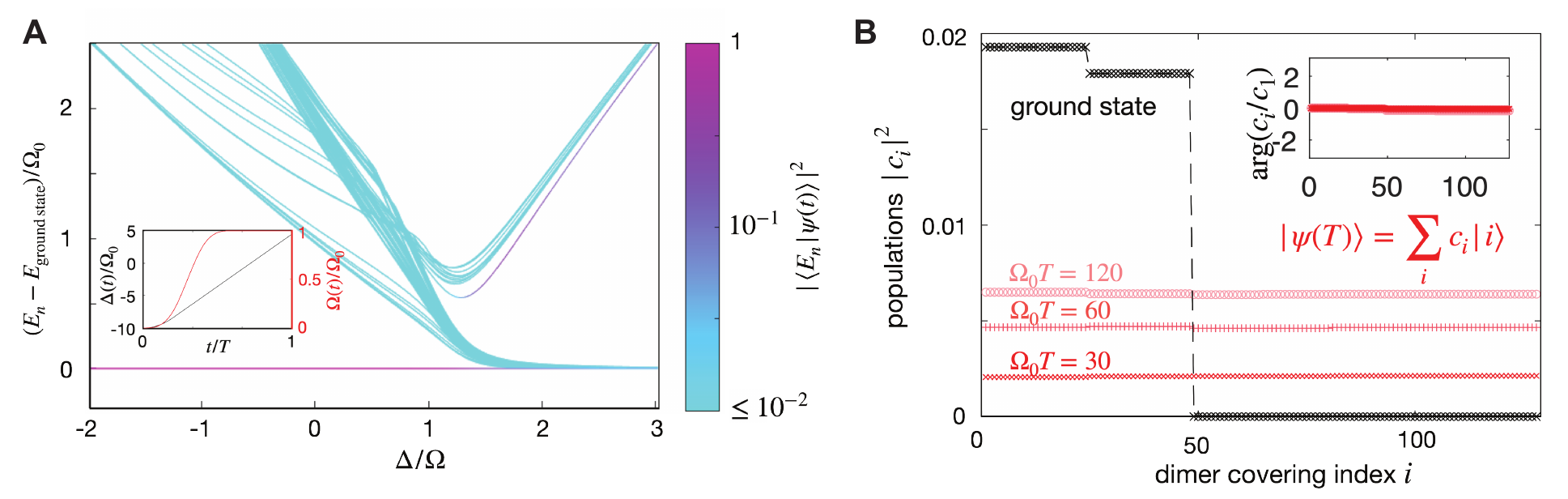}
    \caption{\textbf{Dynamical state preparation in PXP model.} (\textbf{A}) Lowest instantaneous eigenstates of the Hamiltonian in \eqref{eq:PXP} for 36 atoms on a torus. Colors indicate the populations of the state generated in the real time quench dynamics with Hamiltonian parameters given in the inset (data is shown for total sweep time $T=60/\Omega_0$). (\textbf{B}) Decomposition of the ground state and the dynamically generated state at the end of the parameter sweep ($\Delta/\Omega_0=5$) over all dimer covering configurations for various sweep durations $T$. The total population in the dimer covering sector is $\sum_{i\in \mathcal{D}}|c_i|^2=0.27,0.60,0.82$ for $\Omega_0 T=30,60,120$, respectively. For the ground state at $\Delta/\Omega=5$ the population in the dimer covering sector is 0.89. The inset shows the phase of each amplitude. For comparison, the experimental state preparation occurs over $\Omega_0 T = 18$.}
    \label{fig:state_prep_ED}
\end{figure*}

The detuning ramps, $\Delta(t)$, which are employed to generate various states, are motivated by the adiabatic principle. For sufficiently slow ramps, the system follows the instantaneous ground state adiabatically \cite{katoAdiabaticTheoremQuantum1950}. In practice, finite coherence times limit the maximum evolution times, and require faster-than-adiabatic sweeps. This  is expected to induce non-adiabatic processes, in particular close to the critical point, where the finite size gap is minimal \cite{zurekDynamicsQuantumPhase2005,dziarmagaDynamicsQuantumPhase2005a,polkovnikovUniversalAdiabaticDynamics2005}.

\begin{figure*}[b!]
    \centering
    \begin{tikzpicture}
    \node at (0,0) {\includegraphics[scale=0.28]{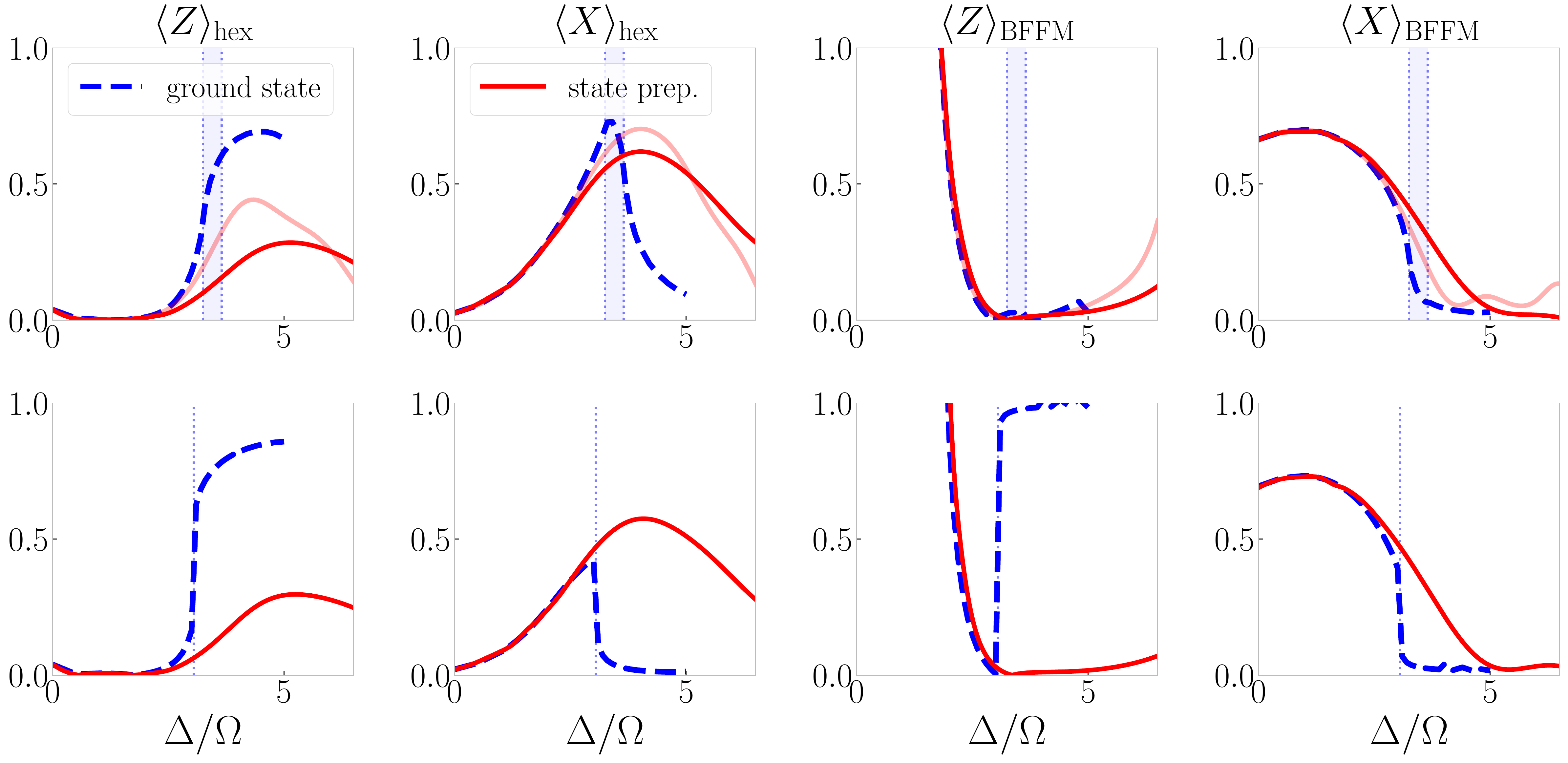}};
    \node at (-8.5,1.75) {$R_\textrm{trunc}=\sqrt{7}a$ $\quad\to$};
    \node at (-8.5,-1.3) {$R_\textrm{trunc}=\sqrt{12}a$ $\quad\to$};
    \end{tikzpicture}
    \caption{\textbf{Dynamical state preparation in the van der Waals model.} Results are for the XC-4 cylinder for $R_b = 2.4a$. The two rows correspond to two different truncation distances, as shown. For each panel, we show both the ground state result (blue dashed, obtained by DMRG) as well as the dynamical state preparation using the protocol in Fig.~\ref{fig:CubicSweep} (red solid, obtained by time-dependent DMRG; lighter solid line is for a sweep at half the speed). For the shorter truncation distance, the ground state hosts a spin liquid (blue shaded region). The diagonal loop around a hexagon is denoted by $\langle Z\rangle_\textrm{hex}$; the off-diagonal loop by $\langle X\rangle_\textrm{hex}$. The BFFM order parameters are evaluated for the open strings that correspond to half of these closed loops. Despite being short strings, due to the small system size, they already show a clear signature of a spin liquid where they both approximately vanish. Note that the ground state data for $R_\textrm{trunc}=\sqrt{7}a$ has a vanishing $\langle Z \rangle_\textrm{BFFM}$, even in the VBS phase: this is a finite-size artefact where the VBS phase consists of local resonances around the circumference. As a  check, we also directly calculated the two-point correlator $\langle n_i n_j \rangle - \langle n_i \rangle \langle n_j \rangle$, which clearly shows the VBS order in the ground state, yet these correlations vanish in the time-evolved states (not shown). We conclude that dynamic state preparation is not sensitive to $R_\textrm{trunc}$, and the resulting state has properties which are similar to those of the ground state spin liquid (at $R_\textrm{trunc}=\sqrt{7}a$) albeit smeared out over a larger region.}
    \label{fig:state_prep_DMRG}
\end{figure*}

To develop an understanding for the quantum many-body states that are generated in such quasi-adiabatic sweeps, we numerically solve the corresponding Sch\"odinger equation to obtain the wavefunction $|\psi(t)\rangle = U(t)|{\psi(0)}\rangle$. We first discuss results from exact numerics on small system sizes of 36 atoms on a torus with $3\times 2$ unit cells, using the simplified PXP-model (eq.~\eqref{eq:PXP}). Fig.~\ref{fig:state_prep_ED}A shows the excitation spectrum of the instantaneous Hamiltonian throughout the sweep. Even though the system size is relatively small, the spectrum distinguishes a disordered region with a unique ground state at $\Delta/\Omega\lesssim 1.5$, and a region whose ground state physics is governed by the dimer covering configurations at $\Delta/\Omega\gtrsim 1.5$. Note that the small system size does not allow to distinguish a spin liquid phase from a VBS phase in this second regime. The color of each individual instantaneous energy eigenvalue in Fig.~\ref{fig:state_prep_ED}A reflects the population of the wavefunction in the corresponding instantaneous eigenstate, $|\langle E_n|\psi(t)\rangle|^2$. We observe that non-adiabatic processes lead to finite population in states with energy $\sim \Delta$ outside the dimer covering subspace. This corresponds to the creation of pairs of monomers, consistent with the experimentally observed generation of a finite density of $e$-anyons. For the sweep profile shown in the inset, the total population in the dimer covering subspace, $\mathcal{D}$, at the end of the sweep is $\sum_{i\in \mathcal{D}}|\langle i|\psi(T)\rangle|^2\approx0.27,\,0.60,\, 0.82$ for total sweep times $\Omega_0 T=30,60,120$ respectively, showing that the defect density can be controlled and reduced by decreasing the sweep rate. 
In Fig. \ref{fig:state_prep_ED}B, we resolve the state $|\psi(t)\rangle$ within $\mathcal{D}$ at the end of the detuning sweep. At this point, the instantaneous ground state consists of a superposition of a subset of dimer covering configurations, akin to a VBS state. Nevertheless, the projection of the dynamically prepared state $|\psi (t) \rangle$ onto $\mathcal{D}$ consists of a superposition of all dimer coverings with nearly equal modulus and phase. This indicates that the system cannot resolve the slow dynamics within the dimer covering subspace during these finite-time sweeps, and instead ``freezes'' into a state that shares the essential features of the spin liquid state. This is consistent with our experimental observation of QSL characteristics in the dynamically prepared states over a relatively large parameter range, without any signatures of a VBS.

To further corroborate this picture, we also performed dynamical DMRG calculations for state preparation in the realistic model with van der Waals ($1/r^6$) interactions using the matrix product operator-based approach developed in Ref.~\cite{Zaletel15}. We consider the infinitely-long XC-4 cylinder. As for the XC-8 results reported above, there is an intermediate spin liquid between the trivial phase and VBS phase for small truncation distance $R_\textrm{trunc} = \sqrt{7}a$: this ground state data is shown as the dashed blue lines in the top row of Fig.~\ref{fig:state_prep_DMRG} (the shaded region highlights the intermediate spin liquid). For larger truncation distance $R_\textrm{trunc}=\sqrt{12}a$, the spin liquid is replaced by a direct first order phase transition (blue dashed lines in bottom row of Fig.~\ref{fig:state_prep_DMRG}). The dynamical state preparation data is shown as a solid red line: dark solid lines corresponds to the same protocol as the experimental data (see Fig.~\ref{fig:CubicSweep}); light solid line is twice as slow as the experiment.

\begin{figure*}
\includegraphics[width=180mm]{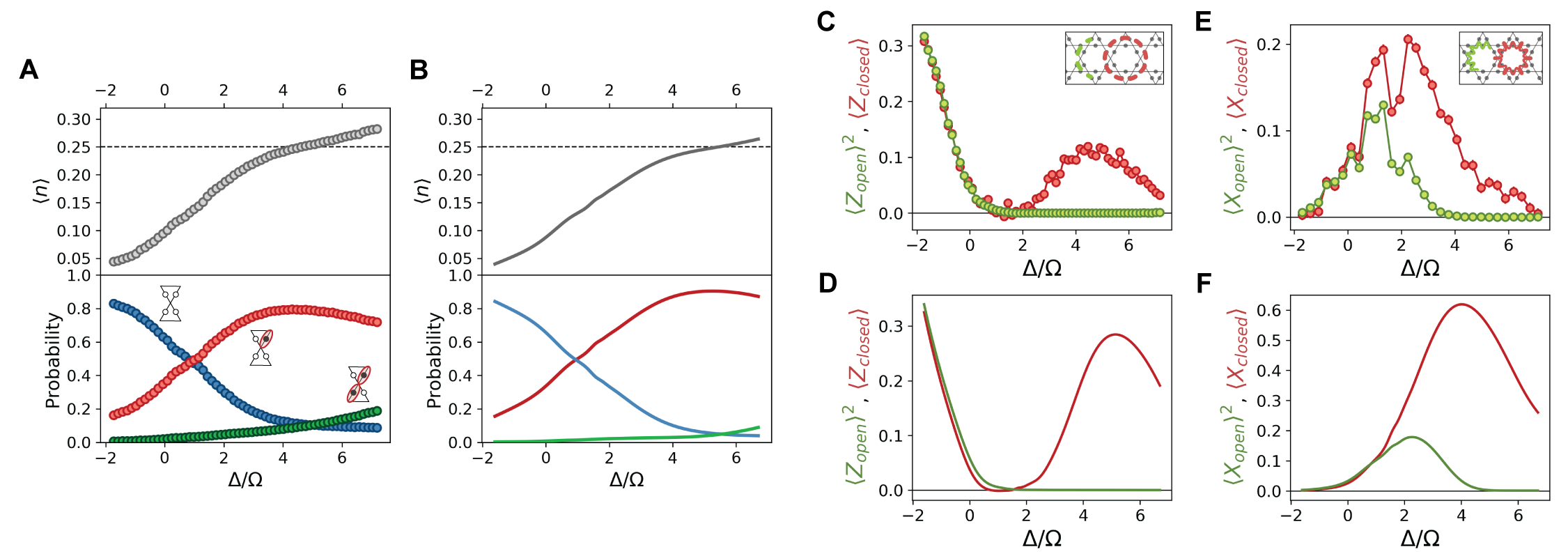}
    \caption{\textbf{Comparison between experimental results and numerical simulations of dynamical state preparation.} The experimental data (\textbf{A},\textbf{C},\textbf{E}) is reproduced from Figs.~1,4 of the main text, while in (\textbf{B},\textbf{D},\textbf{F}) we show the results of the time-dependent DMRG simulations for $R_{trunc}=\sqrt{7}a$, performed on the infinitely-long cylinder with a seven-atom-long circumference (XC-4). }
\label{fig:ExpvsTh}
\end{figure*}

The results in Fig.~\ref{fig:state_prep_DMRG} imply a few salient points. Firstly, as far as dynamical state preparation is concerned, the results for the two truncation distances are very similar: the state preparation seems insensitive to longer-range interactions destroying the intermediate spin liquid in the ground state. Secondly, in both cases, the properties of the time-evolved state are qualitatively very similar to those of the ground state for $R_\textrm{trunc}=\sqrt{7}a$ in the spin liquid regime. The two main differences are: (a) the spin liquid-like state is spread out over a larger region and shifted to the right (minimum of the BFFM order parameters is achieved near $\Delta/\Omega \approx 5$) , and (b) the observables are slightly suppressed compared to their equilibrium values. With regard to the latter, we observe  that the state which was prepared twice as slowly (light red line) gives improved results, in agreement with experimental observations, Fig.~\ref{fig:Z_1GHz}.  This is consistent with the picture that already emerged from the dynamical simulations for the PXP model in Fig.~\ref{fig:state_prep_ED}: even if the ground state is not a spin liquid due to a first order transition to a VBS phase, the dynamically prepared state effectively exhibits spin liquid-like properties, presumably due to the freezing-out of $m$-anyons (which would need to condense to form the VBS phase). Figure \ref{fig:ExpvsTh} demonstrates 
that the results of these dynamical simulations, despite different system sizes and geometry used, are in a  good qualitative agreement with experimental observations. 

\end{document}